\documentstyle[12pt,aasms4]{article}

\def\arcsecpoint{$''\!.$}
\def\deg{$^{\rm o}$}

\lefthead{Kraemer, Crenshaw, Hutchings, Gull, Kaier, Nelson, \& Weistrop}
\righthead{Physical Conditions in the Narrow-Line Region in NGC 4151}

\begin{document}

\title{STIS Longslit Spectroscopy of the Narrow-Line Region of NGC 4151.\\
II. Physical Conditions along Position Angle 221\deg\altaffilmark{1}}

\author{S. B. Kraemer\altaffilmark{2},
D. M. Crenshaw\altaffilmark{2},
J. B. Hutchings\altaffilmark{3},
T. R. Gull\altaffilmark{4},
M. E. Kaiser\altaffilmark{5},
C. H. Nelson\altaffilmark{6},
\& D. Weistrop\altaffilmark{6}
}

\altaffiltext{1}{Based on observations made with the NASA/ESA Hubble Space 
Telescope. STScI is operated by the Association of Universities for Research in 
Astronomy, Inc. under the NASA contract NAS5-26555. }

\altaffiltext{2}{Catholic University of America,
NASA/Goddard Space Flight Center, Code 681,
Greenbelt, MD  20771; stiskraemer@yancey.gsfc.nasa.gov, 
crenshaw@buckeye.gsfc.nasa.gov.}

\altaffiltext{3}{Dominion Astrophysical Observatory, National Research
Council of Canada, 5071 W. Saanich Rd., Victoria, B.C. V8X 4M6, Canada,
John.Hutchings@hia.nrc.ca}

\altaffiltext{4}{NASA's Goddard Space Flight Center, Laboratory for Astronomy
and Solar Physics, Code 681, Greenbelt, MD 20771, gull@sea.gsfc.nasa.gov}

\altaffiltext{5}{Department of Physics and Astronomy, Johns Hopkins University,
Baltimore, MD 21218, kaiser@munin.pha.jhu.edu}

\altaffiltext{6}{Department of Physics, University of Nevada, Las Vegas,
4505 Maryland Parkway, Las Vegas, NV 89154-4002, cnelson@physics.unlv.edu,
weistrop@nevada.edu}

\begin{abstract}

  We have examined the physical conditions in the narrow-line region
of the well-studied Seyfert galaxy NGC 4151, using long-slit spectra
obtained with the {\it Hubble Space Telescope}/Space Telescope Imaging
Spectrograph ({\it HST}/STIS). The data were taken along a position angle of 
221\deg,
centered on the optical nucleus. We have generated photoionization models
for a contiguous set of radial zones, out to 2\arcsecpoint3 in projected position
to the southwest of the nucleus, and 2\arcsecpoint7 to the northeast. 
Given the uncertainties in the reddening correction, the calculated 
line ratios successfully matched nearly all the dereddened ratios. We find 
that the narrow-line region consists of dusty atomic gas photoionized by a 
power-law continuum that has been modified by transmission through a mix of low and high 
ionization gas, specifically UV and X-ray absorbing components. The physical 
characteristics of the absorbers resemble those observed along our line of sight to the nucleus,
although the column density of the X-ray absorber is a factor of ten less than
observed. The 
large inferred covering factor of the absorbing gas is in agreement with the 
results of our previous study of UV absorption in Seyfert 1 galaxies. We 
find evidence, specifically the suppression of L$\alpha$,
that we are observing the back-end of dusty ionized clouds in the region southwest
of the nucleus. Since these clouds are blueshifted, this supports the interpretation of the cloud kinematics
as being due to radial outflow from the nucleus. We find that the 
narrow-line gas at each radial position is inhomogeneous, and can be modeled 
as consisting of a radiation-bounded component and a more tenuous, 
matter-bounded component. The density of the narrow-line gas drops
with increasing radial distance, which confirms our earlier results, and
may be due to the expansion of radially outflowing emission-line clouds.

\end{abstract}


\section{Introduction}

NGC 4151 is one of the nearest (z=0.0033) Seyfert galaxies, and among the most
extensively studied. Based on its strong broad (FWHM $\sim$ 5000 km s$^{-1}$)
permitted lines and our direct line of sight to the non-thermal
continuum source, NGC 4151 is generally classified as a Seyfert 1 galaxy,
although its continuum and broad emission lines are extremely variable and
at one time the broad component nearly disappeared (Penston and Perez 1984), 
at which time
the spectrum of NGC 4151 resembled that of a Seyfert 2 galaxy. It has an
extended emission-line region (i.e. narrow-line region, NLR) which is often 
cited as a an example of the biconical morphology resulting from collimation of 
the 
ionizing radiation (cf. Schmitt \& Kinney 1996). A linear radio structure
(``jet'') extends 3\arcsecpoint5 southwest of the nucleus (Wilson \& Ulvestad 1982)
and may be the result of the interaction of radially accelerated 
energetic particles with the 
interstellar magnetic field of NGC 4151. There have been suggestions, most recently by 
Wilson \& Raymond (1999), that the interaction of the jet with interstellar
gas could be the source of 
the narrow-line emission. 
Much of the gas in the NLR appears to be radially outflowing from the nucleus
(Hutchings et al. 1998), although there is no apparent correlation of the
kinematics of the emission-line gas and the radio structure (Kaiser et al.
1999), which suggests another source for the radial outflow, such as 
a wind emanating from the inner regions of the nucleus (cf. Krolik 
\& Begelman 1984). 

 One of the most interesting spectral characteristics of NGC 4151
is the presence of absorption by atomic gas, seen in the UV (cf. Weymann et al.
1997) and X-ray (cf. Holt et al. 1980). The variability of the absorption
features suggest that the absorbing material must lie close to the nucleus,
within the inner few parsecs for the UV absorber (Espey et al. 1998) and within
a parsec for the X-ray absorber (Yaqoob, Warwick, \& Pounds 1989). 
Although the absorbers
can only be detected along our line of sight to the nucleus, 
their effect on the gas further out should be apparent (Kraemer et al. 1999)
if they have a large covering factor, as suggested by Crenshaw et al. (1999).
In fact, Alexander et al. (1999) have demonstrated that narrow emission
lines of NGC 4151
could be the result of photoionization by an absorbed continuum. However, the 
heterogeneous nature of their data, and the lack of spatially resolved spectra, 
prevented them from constraining the spectral energy distribution (SED)
of the ionizing continuum, other than by noting the apparent paucity
of photons just above the He~II Lyman limit (54.4 eV).

 We have obtained low-dispersion long-slit data, at a position angle
PA $=$ 221\deg, using {\it HST}/STIS. From these data, coupled
with detailed photoionization models, we 
are able to determine the physical conditions of the NLR gas, including any
radial variations. As we will show in the following sections, 
there is no evidence at this position angle for other sources of ionization other than 
photoionization by the nuclear continuum. The NLR does indeed appear to be
ionized by an absorbed continuum, as proposed by Alexander et al. (1999).
Furthermore, based on our modeling, we can place much tighter constraints
on the SED of the ionizing radiation. Finally, these data provide
further evidence that the NLR gas is outflowing radially from the
nucleus of NGC 4151.

\section{Observations and Analysis}

 The details of the observations and data reduction are described in Nelson et al.
(1999, hereafter Paper I). To summarize, we used the G140L grating with the 
far-UV MAMA detector and the G230LB, G430L, and G750L gratings with the
CCD detector, to cover the entire 1150 -- 10,270 \AA~ region, at a
resolving power of $\lambda$/$\Delta\lambda$ $=$ 500 -- 1000.
We used a narrow slit width of 0\arcsecpoint1 to avoid degradation of the spectral
resolution. We extracted fluxes in bins with lengths of 0\arcsecpoint2
in the inner 1$''$ and 0\arcsecpoint4 further out to maximize the
signal-to-noise ratios of the He~II lines (see below) and yet isolate
the emission-line knots identified in our earlier paper (Hutchings et al.
1998).
 
  Our measurement techniques follow those used in previous papers (cf.,
Kraemer, Ruiz, \& Crenshaw 1998). We obtained fluxes for nearly all of the
emission lines by direct integration over a local baseline (see Paper I).
For the blended lines of H$\alpha$ and [N~II] $\lambda\lambda$6548, 6584,
and [S~II] $\lambda\lambda$6716, 6731, we used the [O~III] $\lambda$5007 line
as a template. We determined the reddening from the observed He~II $\lambda$1640/$\lambda$4686
ratios, assuming an intrinsic ratio of 7.2 (expected from NLR temperatures
and densities), and the Galactic reddening curve of Savage \& Mathis (1979).
We determined the errors in the observed ratios (relative to H$\beta$) from 
the sum in quadrature of photon noise and different reasonable continuum 
placements. Errors in the dereddened ratios include an additional contribution
(added in quadrature) from the uncertainty in reddening (propagated from
errors in the He~II ratios); the latter
dominates the errors in the dereddened UV lines.

Tables 1 and 2 give the dereddened ratios and errors in each bin, 
along with the reddening and its uncertainty, and 
the total H$\beta$ flux in that bin (the observed ratios are included in
Paper I). For positions that include 
a knot of emission that we identified in an earlier paper (Hutchings et al. 
1998), we list that knot. We note that the sensitivity of the G230LB grating is 
very low at the short wavelength end, and thus we could not detect the 
N~III] $\lambda$1750 line. The errors in the lines longward of 7000 \AA\ are 
probably underestimates since this region of the G750L spectrum suffers from 
fringing (Plait \& Bohlin 1998).

\section{General Trends in the Line Ratios}

Figure 1 shows various trends in the reddening and line ratios as a function of 
projected distance along the slit. The central position of each bin is plotted 
along the horizontal axis, with the convention that negative numbers represent 
the blueshifted (southwest) side. From the first plot, we see that there is 
significant reddening at many positions, whereas a few positions have 
essentially no reddening. The reddening does not appear to vary in a uniform 
fashion over this region, which suggests that it is not due to an 
external screen, but is associated with the emission-line knots.
If we look at the inner region ($\pm$ 
1$''$), where the errors are smaller, there {\it may} be a trend, in that the 
reddening seems to drop from blueshifted to redshifted side.

The dereddened L$\alpha$/H$\beta$ plot shows a very interesting trend: this 
ratio increases dramatically from the blueshifted to redshifted side. Since
there is no correlation between the reddening and the L$\alpha$/H$\beta$ 
ratio, this trend cannot be 
explained by a systematic error in determining the reddening. In the inner 
region, higher values of reddening on the redshifted side (see the previous 
plot), would only make the L$\alpha$/H$\beta$ trend steeper.

The L$\alpha$/H$\beta$ trend can be explained by radial outflow of 
optically thick clouds that contain dust. In such a cloud, 
resonance-line photons (and L$\alpha$ in particular) are selectively destroyed 
by multiple scatterings and eventual absorption by dust (cf. Cohen, Harrington,
\& Hess 1984). The observer sees a much higher L$\alpha$/H$\beta$ ratio when looking at 
the illuminated face of a cloud (as opposed to the back side), since escape of 
the L$\alpha$ photons is much easier in this direction. Thus, we expect higher
L$\alpha$/H$\beta$ ratios for clouds on the far side of the nucleus, which in 
this case corresponds to redshifted clouds, indicating radial outflow (see
Capriotti, Foltz, \& Byard 1979). Although the UV absorption lines
are blue-shifted (Weymann et al. 1997), indicating that components of 
gas in our line-of-sight are outflowing, the suppression of L$\alpha$ is 
the strongest evidence that the emission line gas in the inner NLR of NGC 4151 
in outflowing radially (for 
further discussion of the NLR kinematics, see Hutchings et al. 1998; 
Kaiser et al. 1999; Paper I).

The [O III]/H$\beta$ plot verifies a trend that we first observed in 
our slitless data (Kaiser et al. 1999). In the inner NLR, this ratio decreases 
with increasing projected distance from the nucleus. Since the ionization parameter is proportional to n$_{H}$$^{-1}$r$^{-2}$, (where 
n$_H$ is the atomic hydrogen density and r is the distance from the nucleus), 
we suggested that n$_H$ decreases roughly as r$^{-1}$ in the inner NLR, and 
more steeply with increasing distance (Kaiser et al. 1999). However, we will
show in the discussion of the model results that the closest radial points
are best fit by assuming a nearly constant ionization parameter, which
indicates that n$_{H}$ decreases more as r$^{-1.6~to~-1.7}$.

In the last plot, we show the H$\gamma$/H$\beta$ ratio as a function of 
projected distance. The ratio of these H recombination lines should show no 
trend, since it only has a slight dependence on temperature and density.
In fact, we see that there is no observed trend, and are therefore confident 
that there are no large systematic errors in the ratios.

\section{Photoionization Models}

The basic methodology which we employ and the details of our photoionization code have 
been described in our previous publications (cf. Kraemer 1985; 
Kraemer et al. 1994). The models assume plane-parallel geometry and
emission-line photon escape out either the illuminated face or the
back-end of the slab. In the latter case, the back surface of the slab was
assumed to be the point where the model was truncated (that is we ignored
any effect from an additional neutral envelope).  In practice,
we used the back-end values for the three positions with a dereddened
L$\alpha$/H$\beta$ $<$ 5. Most of the other line ratios are not affected
by the choice of front- or back-end escape, due to the low column densities 
for these models (see Section 5.1).

 The photoionization models are parameterized in terms of the density
of atomic hydrogen (n$_{H}$) and the dimensionless ionization parameter
at the illuminated face of a cloud:

\begin{equation}
U = {1\over{4\pi~D^2~n_H~c}}~ \int^{\infty}_{\nu_0} ~\frac{L_\nu}{h\nu}~d\nu,
\end{equation}
where L$_{\nu}$ is the frequency dependent luminosity of the ionizing continuum,
D is the distance between the cloud and the ionizing source and 
h$\nu_{0}$ = 13.6 eV. 
     
\subsection{The Ionizing Continuum}

   We have assumed in these simple models that the gas is photoionized
   by radiation from the central AGN. As has often been demonstrated
   (cf., Binette, Robinson, \& Courvoisier 1988), the physical conditions in 
   the NLR, and, hence,
   resulting emission-line spectrum, depend on both the intensity of the radiation
   field and its SED. Along our line-of-sight, the continuum radiation
   from NGC 4151 shows the effects of a complex system of absorbers,
   which is interior to the NLR and
   can alter the SED that the NLR sees drastically. Kraemer et al. (1999) have
   demonstrated that, for typical NLR conditions (U $=$ 10$^{-2.5}$, n$_{H}$ 
   $\leq$ 10$^{5}$ cm$^{-3}$), the ratio of 
   [O~III] $\lambda$4363 to He~II $\lambda$4686 increases as the
   depth of the absorption near the He~II Lyman edge increases.
   From the observed ratios of these lines (see
   Tables 1 and 2), it is clear that [O~III] $\lambda$4363 is relatively
   strong. Therefore, our approach was to estimate the intrinsic (unabsorbed) SED as 
   simply as 
   possible, and then include the effects of absorption to replicate the
   ionization state and electron temperature of the NLR gas.
   We constrained the physical
   characteristics of the absorbers to resemble those observed along our
   line-of-sight. The SED was then held
   fixed as an input to the models. Although Alexander et al. (1999) 
   determined the general characterstics of the SED from emission-line diagnostics, they did not
   attempt to model the affects of realistic absorbers.

   In order to approximate the intrinsic SED, we used the observed fluxes
   at energies
   where the effects of the absorber are not severe, specifically at energies lower than
   the Lyman limit and above several keV. In the 1 -- 4 keV band, the continuum
   radiation from NGC 4151 has been modified by a large absorbing column of
   gas (Barr et al. 1977; Mushotzky, Holt, \& Serlemitsos 1978), which
   may have multiple components (Weaver et al. 1994). For the purposes
   of these models, we assumed that the absorption at 5 keV is negligible and
   that the X-ray continuum down to 1 keV can be approximated by a 
   power-law (F$_{\nu}$ $\propto$ $\nu^{-\alpha}$) with spectral index,
   $\alpha$ $\approx$ 0.5 (Weaver et al. 1994). For the 5 keV flux, we have 
   used the value from {\it ASCA} observations given in Edelson et al. (1996).

   The far-UV continuum of NGC 4151 has been 
   observed by HUT on both Astro-1 (Kriss et al. 1992) and Astro-2 
   (Kriss et al. 1995) missions and
   with the Berkeley spectrometer during the ORFEUS-SPAS II mission 
   (Espey et al. 1998). Between the two HUT observations, the
   continuum flux at 1455 \AA~ increased by a factor of $\sim$
   5 (Kriss et al. 1995), and somewhat more near the Lyman limit (the ORFEUS
   results are similar to those from Astro-2). Note, however, that
   the continuum irradiating any parcel of NLR gas is time averaged over the
   light crossing time to that parcel. Thus, variations in the 
   nuclear continuum flux on time scales of a few years will be
   insignificant, since our measurements are
   sampling at least 30 years of continuum variations. 
   To be conservative, we assumed a reddening corrected flux at the Lyman 
   limit that is 
   the average of the two HUT values. 

   As noted above, the X-ray continuum can be approximated by a 
   hard power-law, the extrapolation of which would lie far below the 
   observed flux near the Lyman limit. Since many Seyfert spectra
   appear to steepen in the soft X-ray (cf., Arnaud et al. 1985;
   Turner et al. 1999), we assume that the X-ray continuum of NGC 4151
   extends to 1 keV, where the index becomes steeper, continuing down to the
   Lyman limit, and flattens to the canonical value ($\alpha$ $=$ 1) below
   13.6 eV.  The ionizing continuum is thus expressed as
   $F_{\nu}$$=$K$\nu^{-\alpha}$, where

\begin{equation}
    \alpha = 1.0, ~h\nu < 13.6~{\rm eV}
\end{equation}
\begin{equation}
    \alpha = 1.4,~ 13.6eV \leq h\nu < 1000~{\rm eV}
\end{equation}
\begin{equation}
    \alpha = 0.5, ~h\nu \geq 1000~{\rm eV}
\end{equation}
   The total luminosity between 13.6 eV and 1 KeV is $\approx$ 1.2 x 10$^{43}$
   erg s$^{-1}$.
   Although based on their models of the 
   extended ($\geq$ 500 pc) emission-line region, Schulz \& Komossa
   (1993) have argued for the existence of a ``Big Blue Bump'' in
   the EUV continuum of NGC 4151, Alexander et al. (1999) have
   demonstrated that the NLR shows no such effects. Furthermore, as we 
   will show, this simple power-law fit provides a sufficient flux in ionizing
   radiation to power the NLR, even after including the effects of absorption 
   near the He~II Lyman limit. Hence, 
   there is no reason to include a blue bump.
  
   There have been several attempts to constrain the physical properties of the
   X-ray absorber using photoionization models (Yaqoob, Warwick, \& Pounds 
   1989; Weaver et al. 1994; Warwick, Done, \& Smith 1995) and, while
   it is likely to consist of multiple components, there is general
   agreement that U is in the range 0.2-1.1, with column density,
   N$_{H}$ $\sim$ 10$^{23}$ cm$^{-2}$, where N$_{H}$ is the sum of both 
   ionized and neutral hydrogen. One effect of the X-ray absorber is to 
   produce deep O~VII and O~VIII absorption edges (739 eV and 871 eV,
   respectively). Absorption of continuum photons at these energies 
   results in a lower electron temperature in the NLR gas, 
   particularly in the partially ionized, X-ray heated envelope behind the H$^{+}$/H$^{0}$
   transition zone. In addition to the large column of X-ray 
   absorbing material, there is a component of lower column, lower ionization 
   gas, which is the
   source of the UV resonance line absorption which has been
   observed with {\it IUE} (Bromage et al. 1985) and {\it HST}
   (Weyman et al. 1997). Kriss (1998) has modeled several different
   kinematic components of the UV absorber with N$_{H}$
   $\sim$ 10$^{19.5}$ cm$^{-2}$ and U $\geq$ 10$^{-3.0}$. Such an
   absorber is optically thick at the He~II Lyman limit 
   and would produce the effects on emission-line diagnostics discussed
   in Kraemer et al. (1999).
   
   For our models, we
   used a single X-ray absorber and a single UV absorber along each of the
   two lines-of-sight (SW and NE). For the
   X-ray absorber we obtained the best fit to the average NLR conditions
   by assuming U $=$ 1.0 and N$_{H}$ $=$
   3.2 x 10$^{22}$ cm$^{-2}$. For the UV absorber, we used U $=$ 10$^{-3.0}$ 
   and N$_{H}$ $=$ 2.8 x 10$^{19}$ cm$^{-2}$ and 5.6 x 10$^{19}$ cm$^{-2}$
   for the southwest (SW) and northeast (NE) sides, respectively. The different
   column densities were chosen to fit the differences in the relative
   strengths of the [O~III] $\lambda$4363 and He~II $\lambda$4686 lines
   on each side of the nucleus. The unabsorbed and transmitted SEDs
   are shown in Figure 2 and 3. Note that we have assumed that the UV absorber
   is external to the X-ray absorber, although the NLR models are 
   not particularly sensitive to the relative locations of the absorbers.

\subsection{Model Input Parameters}

   For the sake of simplicity, we have assumed that 
   the measured distances are the true radial distances on the NE side of the nucleus
  (note that we have assumed H$_{0}$ = 50 km s$^{-1}$ Mpc$^{-1}$ for this
  analysis, which places NGC 4151 at a distance of 20 Mpc, and thus
  1\arcsecpoint0~ corresponds to 100 pc; this was done so that our results could be
  compared more easily with previous studies of the NLR).
  In our model, the smaller column of the UV absorber towards the SW permits
  more ionizing photons to reach the NLR. Since we do not see strong evidence
  for higher ionization in the SW, we have scaled the distances 
  by a factor of 1.2. Interestingly, on the SW side we are seeing some 
  clouds which may have a larger component of velocity in our line-of-sight
  (Kaiser et al. 1999), which indicates a larger projection effect, and
  supports the scaling of observed radial distances.

Although the UV semi-forbidden emission lines from doubly ionized 
carbon, nitrogen, and oxygen can be used to determine the relative abundances
of each element (Netzer 1997), the poor signal-to-noise quality our long-slit 
data between 1650 \AA~  and 1800 \AA~ make reliable measurement of the
strengths of O~III] $\lambda$1663 and N~III] $\lambda$1750 impossible. 
However, the relative strengths of optical forbidden lines show no evidence
of peculiar abundances. Therefore, we have assumed 
roughly solar abundances for these models (cf. Grevesse
\& Anders 1989). The abundances, relative to H by number, are He=0.1,
C=3.4x10$^{-4}$, O=6.8x10$^{-4}$, N=1.2x10$^{-4}$, Ne=1.1x10$^{-4}$,
S=1.5x10$^{-5}$, Si=3.1x10$^{-5}$, Mg=3.3x10$^{-5}$, and Fe=4.0x10$^{-5}$.
However, as we will discuss, there is evidence for 
enhancement of iron close to the nucleus.

 As we discussed in Section 3, the L$\alpha$/H$\beta$ ratio is strong
evidence that there is cosmic dust mixed in with the emission-line gas.
However, both C~II] $\lambda$2328 and Mg~II $\lambda$2800 are relatively
strong lines (about equal to H$\beta$, after correcting for
reddening), which would indicate that a substantial fraction of these
elements remain in gas phase. Therefore, we have assumed
that the amounts of silicate and carbonate dust grains are 50\% 
and 20\% of the Galactic values, respectively, with depletions of elements
onto dust grains, scaled accordingly, as per Seab \& Shull (1983).

 At all of the radial locations sampled, the [O~II] $\lambda$3727 and
[S~II] $\lambda\lambda$6716, 6731 lines are among the strongest optical 
emission lines. Since 
the critical electron density for [O~II] $\lambda$3727 is n$_{e}$ 
$=$ 4.5 x 10$^{3}$ cm$^{-2}$ (de Robertis \& Osterbrock 1984), 
a substantial fraction of the emission-line gas must be at densities n$_{H}$ $\leq$ a few times 
10$^{4}$ cm$^{-3}$, although some of this emission may be due
to projection of low density gas at larger radial distances. The average ratio of 
[S~II] $\lambda$ 6716/$\lambda$6731 is near unity, which is consistent with 
these lines arising in the partially ionized envelope associated with the [O~II] 
clouds, and, hence, where n$_{e}$ $<$ n$_{H}$.

 The spectra at each radial point show emission lines from a wide
range in ionization state. For example, emission from O$^{0}$ through O$^{+3}$
are detected at every location, and, in several cases, N$^{0}$ through N$^{+4}$
are seen. It is 
plausible that the wide range in physical conditions indicated by the
emission-line spectra is the result of local density inhomogeneities in the 
NLR gas. To model this
effect, we have assumed that the gas in which the [O~II] $\lambda$3727 
emission arises is optically thick (radiation-bounded) and the 
principal source of hydrogen 
recombination radiation. To this we added a second component of less
dense, optically thin (matter-bounded)
gas, which is the source of the high ionization lines, such as 
C~IV $\lambda$1550, [Ne~V] $\lambda\lambda$3346, 3426, and most of the
He~II emission. The idea that high excitation lines
arise in a tenuous component distributed through the NLR region of NGC 4151
has been discussed by Korista \& Ferland (1989), although in our models
the matter-bounded component is co-located with and has the
same radial dependence in physical conditions as the denser component.
We determined the relative contributions of these two components by
given approximately equal weight to the predicted vs. observed fits for the 
following line
ratios: [O~III] $\lambda$5007/[O~II]$\lambda$3727, He~II $\lambda$4686/H$\beta$,
and C~IV $\lambda$1550/H$\beta$. The first is sensitive to ionization parameter,
the second to SED and the fraction of matter-bounded to
radiation-bounded gas, and the last is a trace of the high-ionization 
component (although sensitive to the effects of dust). In one case, 
the red-shifted core (0\arcsecpoint1 -- 0\arcsecpoint3 NE),
the C~III] $\lambda$1909/H$\beta$ ratio was included, since the spectrum 
suggests the presence of an additional high density component in which
the oxygen lines might be collisionally suppressed.
Since there are no obvious constraints on the size of 
the matter-bounded component, other than the extraction length along the
slit, we truncated the model such that its physical extent was
approximately 10 times that of its radiation-bounded companion.
We have not considered the case where the high density gas is embedded within
the low density gas, although this is certainly possible. Thus, in our models, 
each component sees the ionizing source directly.

 The input parameters for the models are given in Table 3. As noted above,
the density of the dominant (radiation-bounded) component is $\leq$ 2 x 10$^{4}$ cm$^{-3}$ at the
innermost points, and falls off monotonically with radial distance. We have 
held U roughly constant for this component within the inner 1$''$ 
($\sim$ 100 pcs),
allowing it to drop for the outermost points. The ionization parameter
for the matter-bounded component was allowed to vary to produce the
best fit to the observed emission-line ratios when its contribution
was added to that of the radiation-bounded component, however, its
density is also a decreasing function of radial distance. The component-weighted
model densities as a function of radial distance are plotted in Figure 4,
along with power law fits, showing that density falls off as 
n$_{H}$ $\propto$ r$^{-s}$, where s $=$ $-$1.6 to $-$1.7. The one 
exception to the two-component scheme is position 0\arcsecpoint1 --
0\arcsecpoint3 NE, for which
three components were required, which might be expected since projection 
effects are likely to be greatest at small projected radial distances.

\section{Model Results}

The model results are shown graphically in Figures 5 and 6. As a measure of
the goodness of the fit, we have included dotted lines to mark a factor of two
divergence between the model prediction and the observed line strength.
Given the simplicity of the 
models, the fact that the large majority of the emission lines are well fit
validates our basic assumption, e.g. that photoionization 
is the dominant mechanism for ionizing the 
NLR gas, and that the densities, dust fraction and abundances used for the
models are approximately correct. Also, our estimated central source
luminosity is sufficient to power the NLR, which implies that 
1) there is no evidence for anisotropy of the continuum radiation, and 2)
the average luminosity of the central source has been roughly the
same for several hundreds of years. Furthermore, by fitting lines from ions with 
a wide range of 
ionization potential (e.g., 7.6 eV for Mg~II; 97.0 eV for Ne~V) the 
model predictions confirm that the SED of the ionizing continuum is
correct, including the assumptions regarding the effects of the UV and
X-ray absorbers. 

\subsection{Fit to the Observations}

As expected, we have achieved good fits simultaneously for 
the strengths of C~IV $\lambda$1550, [O~III] $\lambda$5007, [O~II] 
$\lambda$3727,
and He~II $\lambda$4686, relative to H$\beta$. Even though we weighted the 
contributions of the individual components bases on these lines, the
quality of the fit indicates that we have made a good approximation to the
range of physical conditions at each point. Also, the composite models give
good predictions for high ionization lines, such as O~IV] $\lambda$1402 
and [Ne~V] $\lambda$3426, and lower ionization lines, including those
formed primarily in the H$^{+}$ zone, such as [Ne~III] $\lambda\lambda$3869, 
3968 and [N~II] $\lambda\lambda$6548, 6584, and those formed in
the H$^{0}$ zone, such as [O~I] $\lambda\lambda$6300, 6364, [N~I] 
$\lambda\lambda$5198, 5200, and [S~II] $\lambda\lambda$6716, 6731. Not only does the
quality of the fit further validate our simple two component scheme but, also,
the fact that lines from the neutral envelopes are well-fit supports
the assumption that much of the NLR gas is radiation-bounded. 

Our central hypothesis was that the NLR was irradiated by an ionizing 
continuum which was absorbed by intervening gas close to the nucleus.
As noted above, the fit to SED sensitive lines, such as He~II $\lambda$4686,
and lines from a wide range of ionization states indicate that the
overall SED is approximatedly correct. The electron temperature is also
sensitive to assumptions about the SED. For the models of the bins NE
of the nucleus, the predicted ratio of [O~III] $\lambda$4363/[O~III] 
$\lambda$5007 is, on average, within a factor of $\approx$ 1.5 of that observed, which
indicates a reasonably accurate prediction for the electron temperature.
The relative strength of [O~III] $\lambda$4363 is somewhat underpredicted
in the models for the bins SW of the nucleus. We take this as an
indication that we somewhat overestimated the absorption at energies greater 
than 100 eV, which is primarily due to the X-ray absorber (see
Figure 3). Since the strength of [O~I] $\lambda$6300
is sensitive to the electron temperature in the H$^{0}$ zone, 
the undeprediction of [O~I] $\lambda$6300 in the SW is additional evidence that 
we have overestimated of the continuum absorption in the X-ray. Nevertheless, the main point 
is that the relative suppression of 
the He~II line strength in gas with a fairly high state
of excitation indeed results from ionization by a continuum which has a paucity of 
photons near the He~II Lyman limit, as suggested by Alexander et al. (1999)
and Kraemer et al. (1999), and that it appears that the much of the NLR
in NGC 4151 sees the same continuum.

In Figure 7 we compare the observed ratio of [S~II] $\lambda$6716/$\lambda$6731
to that predicted by the models. The observed ratio shows a general increase 
with radial distance, also seen in the model results, indicating 
a fall-off in electron density (Osterbrock 1989). The models 
overpredict the value of the $\lambda$6716/$\lambda$6731 ratio for several 
bins, particularly 
those closest to the nucleus, which is evidence that the actual densities
are greater than those assumed for the models, although the differences
are generally a factor of a few. This may result from our 
assumption of constant density for each component, since it is
certainly plausible that the inner regions of the clouds are somewhat
denser than the hotter, ionized outer parts, whether or not the clouds
are pressure confined. Regarding pressure confinement, the two components
are not in pressure equilibrium with one another, since the densities
generally differ by factors of $\sim$ 30 (see Table 3), while the temperatures 
differ by factors of $\sim$ 2 (see Table 4). 
The simplest explanation is that the clouds
are not fully pressure confined and are expanding as they traverse the NLR,
which may be a natural explanation for the drop in density as a function
of radial distance. 

As noted in Section 3, the dereddened L$\alpha$/H$\beta$ ratios for some
regions show the effects of resonance scattering and
subsequent destruction of the L$\alpha$ photons by dust mixed in with the
emission-line gas. Kraemer \& Harrington (1986) discussed the importance
of this process in the NLR of the Seyfert 2 galaxy Mrk 3, but with the
current data we also gain insight into the geometry of the NLR of NGC 4151. 
As we have discussed, to calculate the effects of looking at the back-ends
of the clouds, we have made the simple assumption that their physical
size is no larger than that defined by our models. Based on
our predictions for the L$\alpha$/H$\beta$ ratio for bins 0\arcsecpoint9 --
0\arcsecpoint1.1 SW, 1\arcsecpoint1 -- 1\arcsecpoint5 SW,
and 1\arcsecpoint9 -- 2\arcsecpoint3 SW (see Figure 5), there is no evidence that these
L$\alpha$ photons have a longer path length prior to escape. However,
back-end escape greatly underpredicts the resonance lines formed in the 
neutral envelope of the radiation-bounded component, specifically Mg~II 
$\lambda$2800 and C~II $\lambda$1335. In each case, the models predictions
of the relative strengths of these lines are within the errors if we
calculated the escape out the front-end of the clouds. Therefore, either the
clouds are not as physically thick as we have assumed or there is
a sufficient velocity gradient across the cloud that our assumptions
about line profile and frequency shift (see Kraemer 1985) are no longer
correct. In any case, based on our results for L$\alpha$, the photon path
length, dust fraction internal to the clouds and viewing aspect that we
have assumed are all approximately correct. Also, there are several other
clouds (bins 0\arcsecpoint5 -- 0\arcsecpoint7 SW, 1\arcsecpoint5 -- 
1\arcsecpoint9 SW, 0\arcsecpoint3 -- 0\arcsecpoint5 NE, and 0\arcsecpoint9
-- 1\arcsecpoint1 NE) for which there is some 
suppression of L$\alpha$ (see Tables 1 and 2), although it is less extreme 
than in the cases we have addressed. This can be attributed to either an
oblique viewing angle through the cloud or a back-end view through a 
smaller column of radiation-bounded gas, although, based on the strengths of 
the collisionally excited lines formed in the neutral envelopes, there is no 
evidence for the latter. As noted, the facts that the suppression of 
L$\alpha$ is much greater in the SW and is due to the effects on internal dust
support our previous conclusions regarding the narrow-line kinematics (Hutchings et al.
1998), specifically that the narrow-line clouds
in the inner $\sim$ 200 pc are in radial outflow from the nucleus.

In Table 3 we list the total column densities for the component models, which
tend to be less than 2 x 10$^{21}$ cm$^{-2}$. The actual column of hydrogen
which line photons traverse while escaping the clouds is typically much less,
since the emission-line are formed throughout the cloud, rather than strictly
at either the surface or deepest point. On the other hand, the average 
reddening over the inner 3$''$, E$_{B-V}$ $\approx$ 0.18, indicates that
we are observing the emission-lines through columns of at least 2.5 x 10$^{21}$
cm$^{-2}$, given the dust fraction assumed and the relationship between
hydrogen column density and reddening determined for the Milky Way (Shull
\& Steenberg 1985). Therefore, the emission-lines are more reddened
than can accounted for by the dust in the clouds in which they are formed
(which is why we compared our model predictions to the dereddened line
ratios in the first place). Instead, we are 
either viewing the clouds through a nonuniform external screen,
since the reddening varies,
or there are multiple dusty clouds along our line-of-sight in many cases,
such that the further ones are reddened by the dust in those nearer.
There is another compelling reason to believe we are seeing superposition
of clouds, which will be discussed in section 5.2.

We have assumed solar abundances for these models and, for all but one of the
regions modeled, there are no discrepancies between the model predictions
and the observations that could be attributable to differences in the
abundances. The one exception is the blue-shifted core 
(0\arcsecpoint1 -- 0\arcsecpoint3 SW), whose
spectrum shows relatively strong lines of [Fe~VII] ($\lambda$6087,
$\lambda$5721, $\lambda$3760, and $\lambda$3588). The model underpredicts
these lines by a factor of $\sim$ 5, although the fit for [Ne~V] 
$\lambda\lambda$ 3426,3346, which should arise in similar conditions,
is good. We achieved a reasonable fit for the [Fe~VII] line by
running a test model in which the iron abundance was increased by
a factor of 2 and no iron was depleted onto grains. Therefore, 
we suggest that the fraction of iron in gas phase is higher in this
region, perhaps due to a liberation of iron from dust grains 
combined with a higher iron abundance. There is evidence for iron
enhancement in the nucleus of NGC 1068 (Netzer \& Turner 1997;
Kraemer et al. 1998), so it is not surprising that a similar effect
is seen in the inner NLR of NGC 4151.

\subsection{Open Issues}

Although the match between the model predictions and the observations are
generally good, there are several apparent discrepancies. First of all,
the predictions for the N~V $\lambda$1240 are generally low. This line
is somewhat suppressed due to the resonance scattering-dust destruction
process discussed previously. If the dust is not uniformly distributed
in the more tenuous components, we might expect to see less suppression of
N~V $\lambda$1240 than the other strong resonance lines, for example
C~IV $\lambda$1550, since the N$^{+4}$ zone
in these models does not fully overlap the C$^{+3}$ zone and is nearer
the illuminated surface of the cloud. Another possible
explanation is that the N~V line is enhanced through scattering of continuum 
radiation by N$^{+4}$ ions (cf. Hamann \& Korista 1996), an effect
which is not included in our photoionization code. This effect 
is particularly important if there is significant turbulence 
in the clouds. Finally, the N~V $\lambda$1240 line may
arise in gas that is not modeled in our simple two-component scheme. For
example, it is possible that there is an even more tenuous, more highly 
ionized component (U $\sim$ 1.0)
filling the narrow-line region. If the narrow-line clouds are driven
outward by a wind originating in the inner nucleus (cf. Krolik \& Begelman
1986), it may be that the N~V emission arises in the wind itself. The models
also underpredict the C~II $\lambda$1335 strength, which is most likely due
to enhancement by resonance scattering of continuum emission,
particularly since the
models predict columns of C$^{+}$ for the radiation-bounded components 
typically $\geq$ 10$^{18}$ cm$^{-2}$ for the clouds within 100 pc of the 
nucleus.
Heckman et al. (1997) have found evidence for this effect in the spectrum
of the Seyfert 2 galaxy I Zw 92 (Mrk 477).
The line may also be affected by the way in which the reddening
correction is determined, as we discuss below.

As can be seen from the ratios of the model/observed line strengths plotted 
in Figures 5 and 6, lines in the near UV band (1700 -- 3000 \AA~) show 
what may be a systematic trend. There are a large number of regions for
which C~III] $\lambda$1909, C~II] $\lambda$2326 and [Ne~IV] $\lambda$2423 
are unpredicted, particularly on the SW side of the nucleus. Since these
lines arise the same physical conditions as several well-fit optical
lines (e.g. [O~III] $\lambda$5007, [N~II] $\lambda\lambda$6548, 6584, etc.)
it is unlikely that their underprediction indicates the presence of
another component of gas. The fact
that the problem is worse in the SW points to the reddening correction as
the cause, since we are observing those clouds through a larger column of 
dust, much of it internal to the emission-line clouds. 
However, the problem appears to be specific to the near UV band, since 
there are no apparent systematic discrepancies in 
the far UV spectra (1200 -- 1700 \AA~). If the strength of the 2200 \AA~ feature
in the interstellar medium of NGC 4151 is less than that in the Milky Way,
we may have overestimated the intrinsic strength of the C~II] $\lambda$2326
and [Ne~IV] $\lambda$2423 lines. Another explanation is required for the 
discrepant C~III] $\lambda$1909 predictions. It is interesting to note that
those regions with the worst fit for the C~III] line (bins 0\arcsecpoint5
-- 0\arcsecpoint7 NE, 0\arcsecpoint3 -- 0\arcsecpoint5 SW,
0\arcsecpoint7 -- 0\arcsecpoint9 SW, and 0\arcsecpoint9 -- 1\arcsecpoint1 SW) have reddening corrected H$\alpha$/H$\beta$ ratios 
somewhat below Case B ($\sim$ 2.9; Osterbrock 1974), which implies that 
these lines have been overcorrected for extinction. The problem may
be the manner in which the reddening is determined, specifically from the 
ratio of He~II $\lambda$1640/He~II $\lambda$4686. We have
assumed that the He~II lines are viewed through the same column of dust
as the other emission lines. This may not be a robust assumption, since most 
of the He~II emission comes from the matter-bounded components. Unfortunately,
the physical sizes and geometry of the emitting regions are 
not well-constrained by our simple models, so it is not possible to 
quantify this effect. In any case, it is likely that the combined
effects of the uncertainty
in the extinction curve in NGC 4151 and different viewing
angles through inhomogeneous emission-line regions account for these
discrepancies.

In Table 4, we list the model predictions for the H$\beta$ flux emitted
at the illuminated face of the cloud. We estimate the size of the emitting 
area, also given in Table 4, by dividing the observed H$\beta$ luminosity
by the emitted flux. For the matter-bounded clouds, this area tends
to be larger than the projected area of the bin, which suggests 1) clouds with
larger depths in the line of sight, or 2) multiple clouds along the line
of sight. We also estimate the depths of the emitting regions by 
dividing these areas by the slit width (10 pc). Within the inner 100 pc,
the depths are
not greater than the apparent radial distance, and are therefore
reasonable in our estimation. 
At larger radial distances, the emitting area for the matter-bounded
component becomes quite large, and would imply that we are seeing the
sum of emission from up to several hundreds of parsecs into the galaxy.
Note, however, that we arbitrarily truncated the size of the matter-bounded
components. Increasing their column densities would result in a
decrease in the size of the required emitting area. Therefore, the model
predictions may be an overestimate of the depth into the galaxy.

One minor problem with the models is that they generally overpredict the
strength of the [S~III] $\lambda$9069 and $\lambda$9532 lines. Osterbrock,
Tran, \& Veilleux (1992) discussed this problem in the context of the
spectral properties of a sample of Seyfert galaxies and suggested that
the observed weakness of these lines may indicate a relative underabundance
of sulfur. However, accurate dielectronic recombination rates have not yet
been calculated for sulfur (Ali et al. 1991) and, as a result, the model
predictions may not be reliable (in fact, it is not clear what rates were used
by Stasinska (1984), whose results were used by Osterbrock et al.). For 
this dataset, the problem with the [S~III] lines may be instrumental. Using
the G750L grating with the STIS CCD at these wavelengths produces an
instrumentally generated interference, or ``fringing'', pattern (Plait \& Bohlin 1998). In our
data, the fringing, coupled with the method for subtracting the 
scattered nuclear spectrum, appears to have caused a certain amount of
destructive interference: we estimate that the observed strengths of the
[S~III] lines have been underestimated by as much as a factor of two
at some locations. Given this, we do not see that
overprediction of the lines is a problem. The fringing effects can be corrected
using a contemporaneous CCD flat-field exposure; we will have this in
our subsequent STIS observations of NGC 4151, and will re-evaluate the
[S~III] problem at that time.

\section{Implications for the Absorbers}

We have shown that the NLR of NGC 4151, at least along one position angle, is 
ionized by an absorbed continuum, and, thus, there are important implications regarding
the distribution of gas in the inner nuclear regions of the galaxy.
We should 
note that we not have placed stringent constraints on the location of the absorber. 
Specifically, we have not yet discussed the possibility that the absorption 
could be due to 
the narrow-line clouds themselves. However, there are good reasons to reject 
such a model. We obtained a good fit to the emission-line ratios
at each radial point assuming that much of the gas is optically thick
to the ionizing radiation, at both the He~II and hydrogen Lyman limits,
unlike the gas that is producing the UV absorption, which is optically thick
only at the He~II Lyman limit. The one location where there may be evidence for a component with
physical conditions similar to the UV absorber is the red-shifted core
(0\arcsecpoint1 -- 0\arcsecpoint3 NE),
specifically the high density, matter-bounded component (see Table 3).
However, at the radial distance of this component ($\geq$ 4 pc from the 
nucleus) the covering
factor is $<<$ 0.01, assuming an ionization cone with an opening
angle of 75\deg~ (Evans et al. 1993), so there could be no
effect on clouds further out. It is possible that the X-ray absorber
extends further into the NLR, and we have argued that there may be a
component of high ionization, tenuous gas that contributes some of the
N~V emission. However, we estimate from U and the size of the
emitting regions that the column densities of such a component 
could not be much greater than a few x 10$^{21}$ cm$^{-2}$, and, therefore,
insufficient to produce the depth of the absorption features in the X-ray
required for our models. Furthermore, we do not see any trends with
radial distance, such as weaker neutral lines or [O~III] $\lambda$4363
emission, that would indicate a drop in electron temperature, and thus an 
increasing column of X-ray absorber.
Thus, it is most likely that both the UV and X-ray absorbers lie within
a few parsecs of the nucleus, as is apparently the case for the absorbers
along our line-of-sight.

Based on the fraction of 
Seyferts found to possess UV and X-ray absorption, the absorber must
have a covering factor between 0.5 -- 1.0 (Crenshaw et al. 1999).
Our analysis
suggests that, along the position angle of these data, the absorber covers the
lines of sight to the NLR, which is in agreement with the average NLR
conditions in NGC 4151 discussed by Alexander et al. (1999).
Unfortunately, we cannot determine the
distribution of NLR gas into the plane of the galaxy from the current dataset,
as we noted in the Section 5.1, although there are likely to be large 
projection effects. 
We have medium resolution
($\lambda$/$\Delta\lambda$ $\approx$ 10,000) slitted observations
planned with {\it HST}/STIS which may provide better constraints on the
distribution of the narrow-line gas and help resolve this question.

\section{Conclusions}

We have examined the physical conditions, along PA 221\deg, in the 
narrow-line gas in the inner $\sim$ 250 pc of the Seyfert 1 galaxy NGC 4151,
using low resolution, long slit data obtained with {\it HST}/STIS.
We have measured the emission-line fluxes at contiguous radial
points along the slit, performed reddening corrections based on the
He~II $\lambda$1640/$\lambda$4686 ratio, and compared the results with
the predictions of detailed photoionization models. The main
results are as follows:

1. The conditions in the NLR gas are the result of photoionization by
   continuum radiation emitted by the central active nucleus. If
   additional excitation mechanisms are present, such as shocks or
   interaction between the radio jet and the interstellar gas, the
   effects must be quite subtle, since the heating and ionization
   of the gas can be successfully modeled by photoionization effects.
   Note, however, that this position angle does not lie directly along
   the radio jet (see Kaiser et al. 1999, and references therein).

2. It is clear that we
   are also seeing the effects of dust imbedded in the clouds.
   Specifically, the L$\alpha$/H$\beta$ ratio on the SW side
   of the nucleus is much lower than predicted by the combination of
   recombination and collisional excitation, and is almost 
   certainly due to line transfer effects within dusty gas. The simplest
   interpretation is that the clouds on the SW are viewed out the
   back-end, rather than from the illuminated face. We used the
   photoionization models to explore this effect and have demonstrated
   that the column densities and dust fraction that we assumed can
   produce the observed supression of L$\alpha$. This is further proof
   that the clouds in the inner NLR are outflowing radially from 
   nucleus, with the SW side approaching us and the NE side receeding from
   us, as discussed our previous papers (Hutchings et al. 1998; Kaiser et
   al. 1999). 

   Although our model predictions indicate that the fraction of dust
   within the emission-line gas is roughly constant across the NLR,
   the amount of reddening varies. Although this could be due to variations
   in the columns density of an external dust screen, it can also be 
   the result of stacking of NLR clouds along our line-of-sight.
   We prefer the latter explanation, since there is evidence, based on 
   the predicted H$\beta$ fluxes, that we are seeing the superposition of 
   clouds.

   3. We have confirmed the earlier results of Kraemer et al. (1999) and
   Alexander et al. (1999), which indicated that the ionizing continuum which 
   irradiates the NLR in NGC 4151 has been absorbed by an intervening
   layer of gas close to the nucleus. With these models we have
   put much more realistic constraints on the SED of the transmitted
   continuum. The fact that the NLR sees an absorbed continuum implies that the covering
   factor of the absorbing material is large, as Crenshaw et al. (1999)
   have suggested. 

\acknowledgments

  S.B.K. and D.M.C acknowledge support from NASA grant NAG 5-4103.

\clearpage

\begin{deluxetable}{lccccc}
\tablenum{3}
\tablecolumns{6}
\scriptsize
\tablecaption{Model Parameters}
\tablewidth{0pt}
\tablehead{
\colhead{Bin} & \colhead{U} &
\colhead{hydrogen density } & 
\colhead{column density } &
\colhead{fraction of } &
\colhead{notes$^{a}$} \\
\colhead{} & \colhead{} &
\colhead{(cm$^{-3}$)} & 
\colhead{(cm$^{-2}$)} &
\colhead{H$\beta$ emission} &
\colhead{} 
}
\startdata
0\arcsecpoint1 -- 0\arcsecpoint3~SW~ & 10$^{-2.63}$ & 1.8 x 10$^{4}$ 
& 1.3 x 10$^{21}$ & 0.70 & RB\\
~~~~~~~~~~ & 10$^{-1.12}$ & 6.6 x 10$^{2}$ & 4.1 x 10$^{20}$ & 0.30 & 
MB\\
0\arcsecpoint3 -- 0\arcsecpoint5~SW~  & 10$^{-2.63}$ & 1.2 x 10$^{4}$
& 1.3 x 10$^{21}$ & 0.70  & RB\\
~~~~~~~~~~ & 10$^{-1.12}$ & 3.7 x 10$^{2}$ & 5.3 x 10$^{20}$ & 0.30
& MB\\
0\arcsecpoint5 -- 0\arcsecpoint7~SW~ & 10$^{-2.57}$ & 5.0 x 10$^{3}$
& 1.3 x 10$^{21}$ & 0.80 & RB\\
~~~~~~~~~~ & 10$^{-1.04}$ & 1.5 x 10$^{2}$ & 5.6 x 10$^{20}$ & 0.20
& MB\\
0\arcsecpoint7 -- 0\arcsecpoint9~SW~ & 10$^{-2.64}$ & 3.0 x 10$^{3}$
& 1.1 x 10$^{21}$ & 0.80 & RB\\
~~~~~~~~~~ & 10$^{-1.16}$ & 1.0 x 10$^{2}$ & 3.7 x 10$^{20}$ & 0.20
& MB\\
0\arcsecpoint9 -- 1\arcsecpoint1~SW~ & 10$^{-2.66}$ & 2.0 x 10$^{3}$
& 1.0 x 10$^{21}$ & 0.80 & RB\\
~~~~~~~~~~ & 10$^{-1.3}$ & 6.0 x 10$^{1}$ & 3.0 x 10$^{20}$ & 0.20
& MB\\
1\arcsecpoint1 -- 1\arcsecpoint5~SW~ & 10$^{-2.66}$ & 1.2 x 10$^{3}$
& 9.3 x 10$^{20}$ & 0.90 & RB\\
~~~~~~~~~~ & 10$^{-1.13}$ & 3.5 x 10$^{1}$ & 2.8 x 10$^{20}$ & 0.10
& MB\\
1\arcsecpoint5 -- 1\arcsecpoint9~SW~ & 10$^{-2.82}$ & 1.0 x 10$^{3}$
& 7.2 x 10$^{20}$ & 0.80 & RB\\
~~~~~~~~~~ & 10$^{-1.13}$ & 2.0 x 10$^{1}$ & 9.6 x 10$^{19}$ & 0.20
& MB\\
1\arcsecpoint9 -- 2\arcsecpoint3~SW~ & 10$^{-2.87}$ & 7.5 x 10$^{2}$
& 6.3 x 10$^{20}$ & 0.85 & RB\\
~~~~~~~~~~ & 10$^{-1.15}$ & 1.5 x 10$^{1}$ & 1.3 x 10$^{20}$ & 0.15
& MB\\
0\arcsecpoint1 -- 0\arcsecpoint3~NE~ & 10$^{-2.67}$ & 1.2 x 10$^{4}$ 
& 1.6 x 10$^{21}$ & 0.50 & RB\\
~~~~~~~~~~ & 10$^{-3.0}$ & 1.0 x 10$^{7}$ & 5.6 x 10$^{19}$ & 0.25 & 
MB\\
~~~~~~~~~~ & 10$^{-1.08}$ & 1.0 x 10$^{5}$ & 5.6 x 10$^{20}$ & 0.25 & 
MB\\
0\arcsecpoint3 -- 0\arcsecpoint5~NE~ & 10$^{-2.67}$ & 1.2 x 10$^{4}$ 
& 1.6 x 10$^{21}$ & 0.90 & RB\\
~~~~~~~~~~ & 10$^{-1.36}$ & 6.0 x 10$^{2}$ & 5.3 x 10$^{20}$ & 0.10 & 
MB\\
0\arcsecpoint5 -- 0\arcsecpoint7~NE~ & 10$^{-2.61}$ & 5.0 x 10$^{3}$ 
& 1.9 x 10$^{21}$ & 0.85 & RB\\
~~~~~~~~~~ & 10$^{-1.08}$ & 1.5 x 10$^{2}$ & 5.6 x 10$^{20}$ & 0.15 & 
MB\\
0\arcsecpoint7 -- 0\arcsecpoint9~NE~ & 10$^{-2.69}$ & 3.0 x 10$^{3}$ 
& 1.3 x 10$^{21}$ & 0.80 & RB\\
~~~~~~~~~~ & 10$^{-1.2}$ & 1.0 x 10$^{2}$ & 4.4 x 10$^{20}$ & 0.20 & 
MB\\
0\arcsecpoint9 -- 1\arcsecpoint1~NE~ & 10$^{-2.69}$ & 2.0 x 10$^{3}$ 
& 1.2 x 10$^{21}$ & 0.90 & RB\\
~~~~~~~~~~ & 10$^{-1.17}$ & 6.0 x 10$^{1}$ & 3.3 x 10$^{20}$ & 0.10 & 
MB\\
1\arcsecpoint5 -- 1\arcsecpoint9~NE~ & 10$^{-3.0}$ & 1.3 x 10$^{3}$ 
& 7.9 x 10$^{20}$ & 0.80 & RB\\
~~~~~~~~~~ & 10$^{-1.16}$ & 2.0 x 10$^{1}$ & 1.2 x 10$^{19}$ & 0.20 & 
MB\\
1\arcsecpoint9 -- 2\arcsecpoint3~NE~ & 10$^{-2.90}$ & 7.5 x 10$^{2}$ 
& 7.7 x 10$^{20}$ & 0.67 & RB\\
~~~~~~~~~~ & 10$^{-1.3}$ & 2.0 x 10$^{1}$ & 2.1 x 10$^{20}$ & 0.33 & 
MB\\
2\arcsecpoint3 -- 2\arcsecpoint7~NE~ & 10$^{-2.90}$ & 5.3 x 10$^{2}$ 
& 7.5 x 10$^{20}$ & 0.75 & RB\\
~~~~~~~~~~ & 10$^{-1.36}$ & 1.5 x 10$^{1}$ & 2.2 x 10$^{20}$ & 0.25 & 
MB\\
\tablenotetext{a}{``RB'' $=$ radiation-bounded; ``MB'' $=$ matter-bounded.}
\enddata
\end{deluxetable}

\clearpage
\begin{deluxetable}{lccccccc}
\tablenum{4}
\tablecolumns{8}
\scriptsize
\tablecaption{Selected Model Output Parameters}
\tablewidth{0pt}
\tablehead{
\colhead{Bin} & \colhead{U} &
\colhead{T$_{initial}$$^{a}$ } &
\colhead{T$_{final}$$^{b}$ } &
\colhead{H$\beta$ flux$^{c}$} &
\colhead{area$^{d}$ }  &
\colhead{distance (pc)} &
\colhead{distance into plane (pc)$^{e}$}
}
\startdata
0\arcsecpoint1 -- 0\arcsecpoint3~SW~ & 10$^{-2.63}$ & 11,750 & --
& 5.09 x 10$^{-1}$ & 8.2$^{2}$ & 23  & 7 \\ 
~~~~~~~~~~ & 10$^{-1.12}$ & 21,450 & 18,100 & 1.88 x 10$^{-2}$ & 28.0$^{2}$
& 23 & 78\\
0\arcsecpoint3 -- 0\arcsecpoint5~SW~ & 10$^{-2.63}$ & 11,650 & --
& 3.39 x 10$^{-1}$ & 6.7$^{2}$ & 47 & 5 \\
~~~~~~~~~~ & 10$^{-1.12}$ & 21,400 & 16,900 & 1.62 x 10$^{-2}$
& 20.0$^{2}$ & 47 & 40 \\
0\arcsecpoint5 -- 0\arcsecpoint7~SW~ & 10$^{-2.57}$ & 11,700 & --
& 1.61 x 10$^{-1}$ & 9.5$^{2}$ & 67 & 9 \\
~~~~~~~~~~ & 10$^{-1.04}$ & 22,200 & 17,900 & 6.6 x 10$^{-3}$ & 23.6$^{2}$
& 67 & 56 \\
0\arcsecpoint7 -- 0\arcsecpoint9~SW~ & 10$^{-2.64}$ & 11,200 & --
& 8.39 x 10$^{-2}$ & 10.6$^{2}$ & 94 & 11 \\
~~~~~~~~~~ & 10$^{-1.16}$ & 20,800 & 17,600 & 3.08 x 10$^{-3}$ & 27.6$^{2}$
& 94 & 76 \\
0\arcsecpoint9 -- 1\arcsecpoint1~SW~ & 10$^{-2.66}$ & 11,000 & --
& 5.43 x 10$^{-2}$ & 18.3 $^{2}$ & 118 & 33 \\
~~~~~~~~~~ & 10$^{-1.3}$~ & 21,200 & 18,600 & 1.44 x 10$^{-3}$ & 56.1$^{2}$
& 118 & 314 \\
1\arcsecpoint1 -- 1\arcsecpoint5~SW~ & 10$^{-2.66}$ & 10,800 & --
& 3.22 x 10$^{-2}$ & 16.3$^{2}$ & 153 & 27 \\
~~~~~~~~~~ & 10$^{-1.13}$ & 21,200 & 18,600 & 7.96 x 10$^{-4}$ & 34.6$^{2}$
& 153 & 120 \\ 
1\arcsecpoint5 -- 1\arcsecpoint9~SW~ & 10$^{-2.82}$ & 10,000 & --
& 1.90 x 10$^{-2}$ & 22.5$^{2}$ & 202 & 23 \\
~~~~~~~~~~ & 10$^{-1.13}$ & 21,200 & 19,900 & 1.53 x 10$^{-4}$ & 125.4$^{2}$
& 202 & 769 \\
1\arcsecpoint9 -- 2\arcsecpoint3~SW~ & 10$^{-2.87}$ &  9,800 & --
& 1.30 x 10$^{-2}$ & 18.8$^{2}$ & 245 & 35 \\
~~~~~~~~~~ & 10$^{-1.15}$ & 20,700 & 19,100 & 1.58 x 10$^{-4}$ & 71.5$^{2}$
& 245 & 511 \\
\tablebreak
0\arcsecpoint1 -- 0\arcsecpoint3~NE~ & 10$^{-2.67}$ & 11,900 & --
& 3.18 x 10$^{-1}$ & 7.0$^{2}$ & 38 & 5 \\
~~~~~~~~~~ & 10$^{-3.0}$~ & 13,900 & 12,800 & 5.90 x 10$^{2}$ & 0.1$^{2}$
& 4 & .001 \\
~~~~~~~~~~ & 10$^{-1.08}$ & 22,300 & 20,500 & 4.26 x 10$^{1}$ & 0.4$^{2}$
& 4 & .02 \\
0\arcsecpoint3 -- 0\arcsecpoint5~NE~ & 10$^{-2.67}$ & 11,900 & --
& 3.18 x 10$^{-1}$ & 8.5$^{2}$ & 38 & 7 \\
~~~~~~~~~~ & 10$^{-1.36}$ & 22,800 & 17,600 & 1.55 x 10$^{-2}$ & 12.8$^{2}$
& 38 & 16 \\
0\arcsecpoint5 -- 0\arcsecpoint7~NE~ & 10$^{-2.61}$ & 12,000 & --
& 1.51 x 10$^{-1}$ & 7.1$^{2}$ & 55 & 5 \\
~~~~~~~~~~ & 10$^{-1.08}$ & 23,600 & 18,700 & 6.29 x 10$^{-3}$ & 14.6$^{2}$
& 55 & 21 \\
0\arcsecpoint7 -- 0\arcsecpoint9~NE~ & 10$^{-2.69}$ & 11,500 & --
& 5.08 x 10$^{-2}$ & 15.9$^{2}$ & 77 & 25 \\
~~~~~~~~~~ & 10$^{-1.2}$~ & 22,100 & 17,600 & 3.56 x 10$^{-3}$ & 30.0$^{2}$
& 77 & 90 \\
0\arcsecpoint9 -- 1\arcsecpoint1~NE~ & 10$^{-2.69}$ & 11,350 & --
& 5.08 x 10$^{-2}$ & 17.2$^{2}$ & 96 & 30 \\
~~~~~~~~~~ & 10$^{-1.17}$ & 22,500 & 19,100 & 1.6 x 10$^{-3}$ & 32.2$^{2}$
& 96 & 104 \\
1\arcsecpoint5 -- 1\arcsecpoint9~NE~ & 10$^{-3.0}$~ & 10,300 & --
& 1.78 x 10$^{-2}$ & 10.9$^{2}$ & 165 & 12 \\
~~~~~~~~~~ & 10$^{-1.16}$ & 22,600 & 20,700 & 2.0 x 10$^{-4}$ & 51.7$^{2}$
& 165 & 267 \\
1\arcsecpoint9 -- 2\arcsecpoint3~NE~ & 10$^{-2.90}$ & 10,300 & --
& 1.21 x 10$^{-2}$ & 14.5$^{2}$ & 200 & 21 \\
~~~~~~~~~~ & 10$^{-1.3}$~ & 22,000 & 20,000 & 3.50 x 10$^{-4}$ & 60.0$^{2}$
& 200 & 360 \\
2\arcsecpoint3 -- 2\arcsecpoint7~NE~ & 10$^{-2.90}$ & 10,200 & --
& 8.39 x 10$^{-3}$ & 17.6$^{2}$ & 240 & 31 \\
~~~~~~~~~~ & 10$^{-1.36}$ & 20,100 & 17,300 & 2.83 x 10$^{-4}$ & 55.3$^{2}$
& 240 & 306 \\
\tablenotetext{a}{electron temperature at the illuminated face of cloud.}
\tablenotetext{b}{electron temperature at the point the model was truncated
(matter-bounded models only).}
\tablenotetext{c}{H$\beta$ flux at illuminated face (ergs sec$^{-1}$ cm$^{-2}$).}
\tablenotetext{d}{surface area (in pc$^{2}$) required to match observed
H$\beta$ luminosity.}
\tablenotetext{e}{emitting area divided by projected slit width (10 pc).}
\enddata
\end{deluxetable}

\clearpage

\figcaption[fig1.ps]{Plots of reddening and various dereddened line ratios
as a function of projected position from the nucleus. Negative values
correspond to the southwest (blueshifted) region.
}\label{fig1}

\figcaption[fig2.eps]{The ionizing continuum assumed for the
SW models. The solid line is the intrinsic continuum (see section 4.1).
The dotted line shows the effects of transmission through the X-ray absorber
(U $=$ 1.0, N$_{H}$ = 3.2 x 10$^{22}$ cm$^{-2}$). The dashed the line
shows the combined effects of the X-ray and UV (U $=$ 0.001, 
N$_{H}$ = 2.8 x 10$^{19}$ cm$^{-2}$) absorbers.
}\label{fig2} 

\figcaption[fig2.eps]{The ionizing continuum assumed for the
NE models. The solid line is the intrinsic continuum (see section 4.1).
The dotted line shows the effects of transmission through the X-ray absorber
(U $=$ 1.0, N$_{H}$ = 3.2 x 10$^{22}$ cm$^{-2}$). The dashed the line
shows the combined effects of the X-ray and UV (U $=$ 0.001, 
N$_{H}$ = 5.6 x 10$^{19}$ cm$^{-2}$) absorbers.
}\label{fig3} 

\figcaption[fig4.eps]{Weighted density for the two-component models
as a function of projected position from the nucleus. The dotted
lines are power law fits such that n$_{H}$ $\propto$ r$^{-s}$.
Negative values correspond to the southwest (blueshifted) region.
}\label{fig4} 

\figcaption[fig5.eps]{The ratios of the model predictions to the
observed and dereddened line strengths for the regions southwest of the 
nucleus. The dashed line indicates a 1:1 correspondence between the 
predicted and observed values; the dotted lines 
indicate a factor of two difference. The values for [N~I] $\lambda$5200 and [S~II] $\lambda$6723 refer
to the combined fluxes of those doublets. The value for C~III] $\lambda$1909 
includes Si~III] $\lambda$1892, and C~II] $\lambda$2326 includes [O~III]
$\lambda\lambda$2321, 2332.
}\label{fig5} 

\figcaption[fig6.eps]{The ratios of the model predictions to the
observed and dereddened line strengths for the regions northeast of the 
nucleus (same format as Figure 5).
}\label{fig6} 

\figcaption[fig7.ps]{Comparison of model predictions (x's) and
observed and dereddened values ($+$'s) of the ratio of 
[S~II] $\lambda$6716/$\lambda$6731 as
a function of projected position from the nucleus. Negative values
correspond to the southwest (blueshifted) region. Corresponding
electron densities range from 4 x 10$^{3}$ cm$^{-3}$ to
1.8 x 10$^{2}$ cm$^{-3}$ ($\lambda$6716/$\lambda$6731 $=$ 0.60 to 1.25) 
(Osterbrock 1989).
}\label{fig7}

\clearpage

\clearpage
\plotone{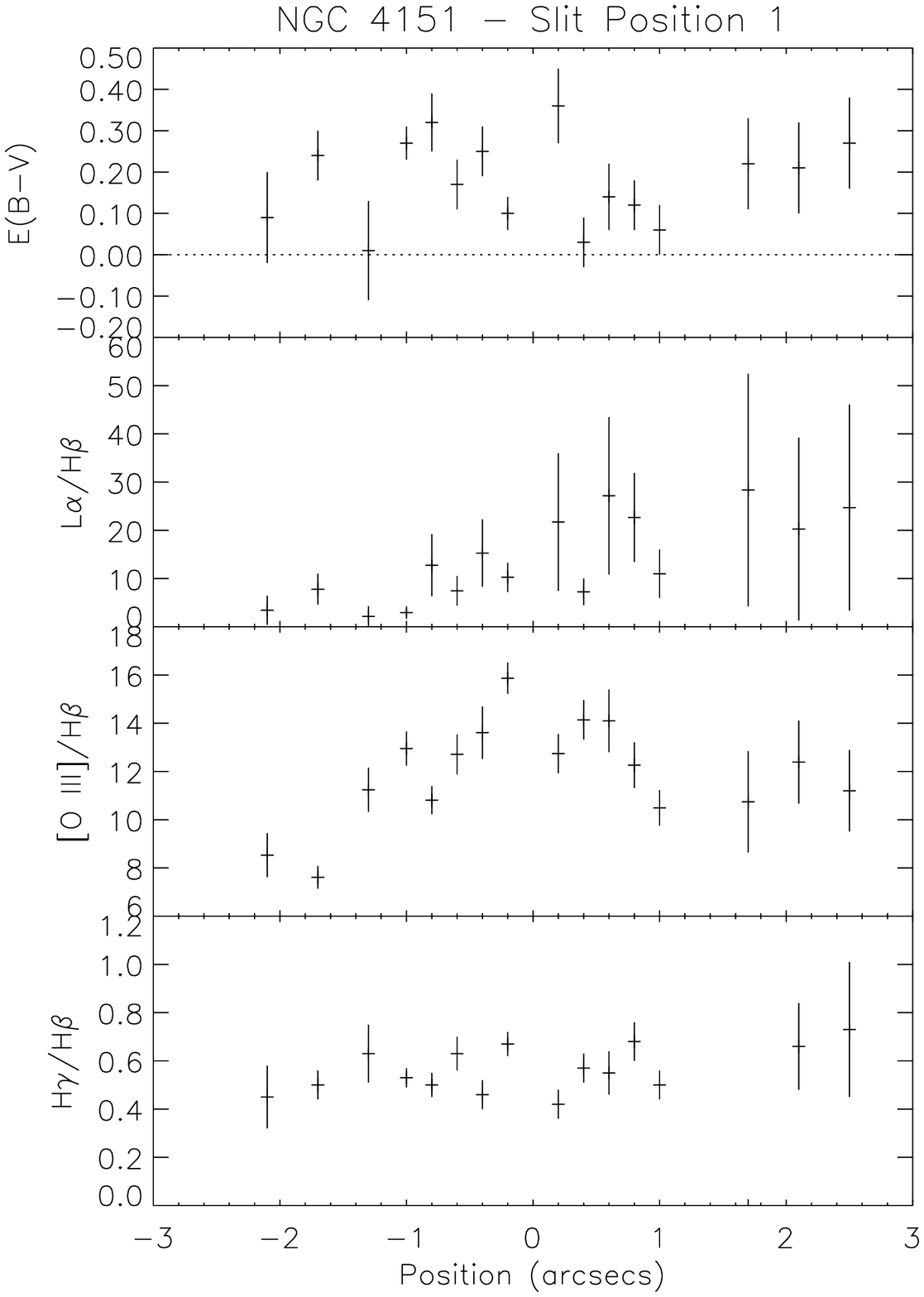}

\clearpage
\plotone{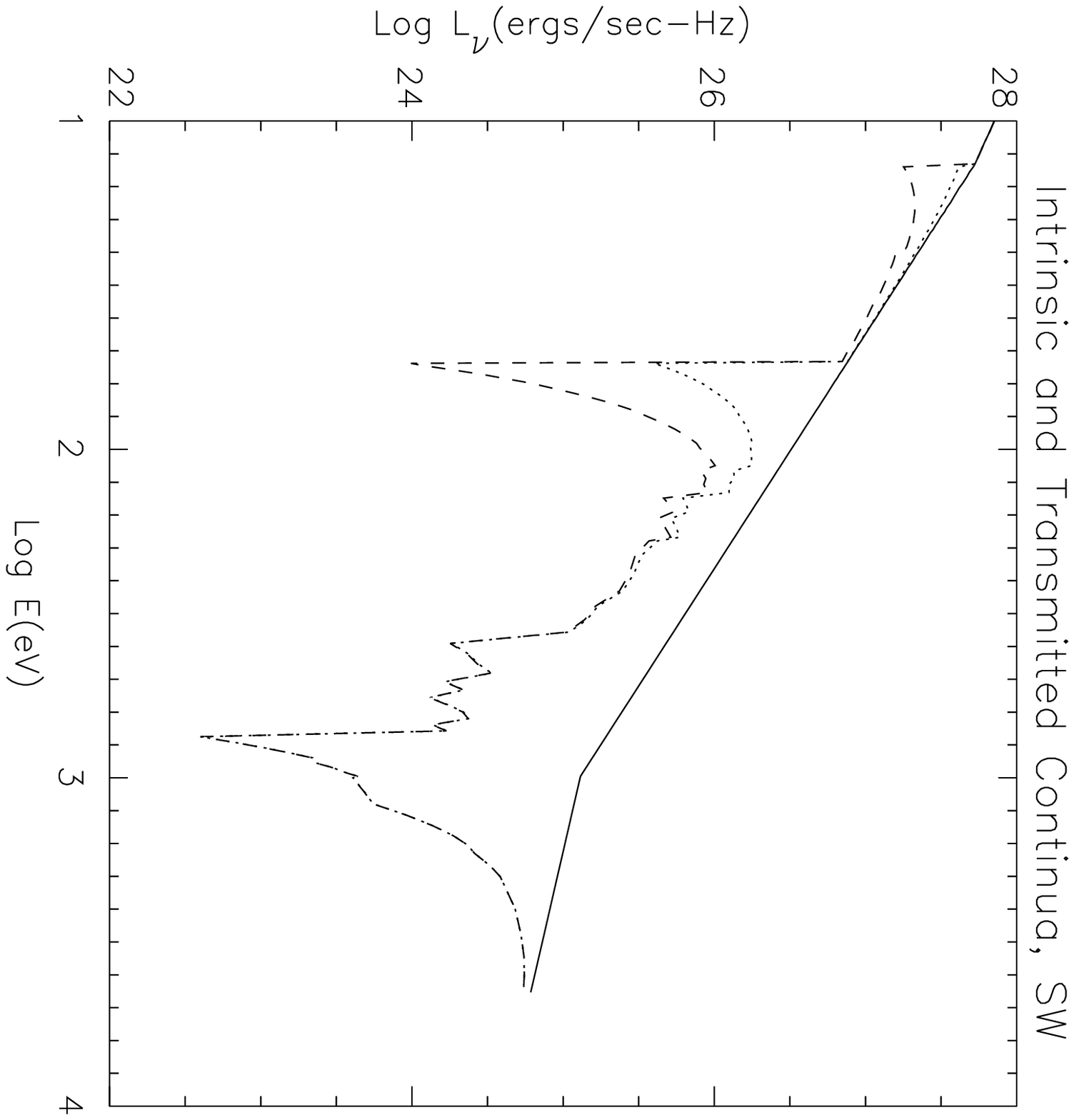}

\clearpage
\plotone{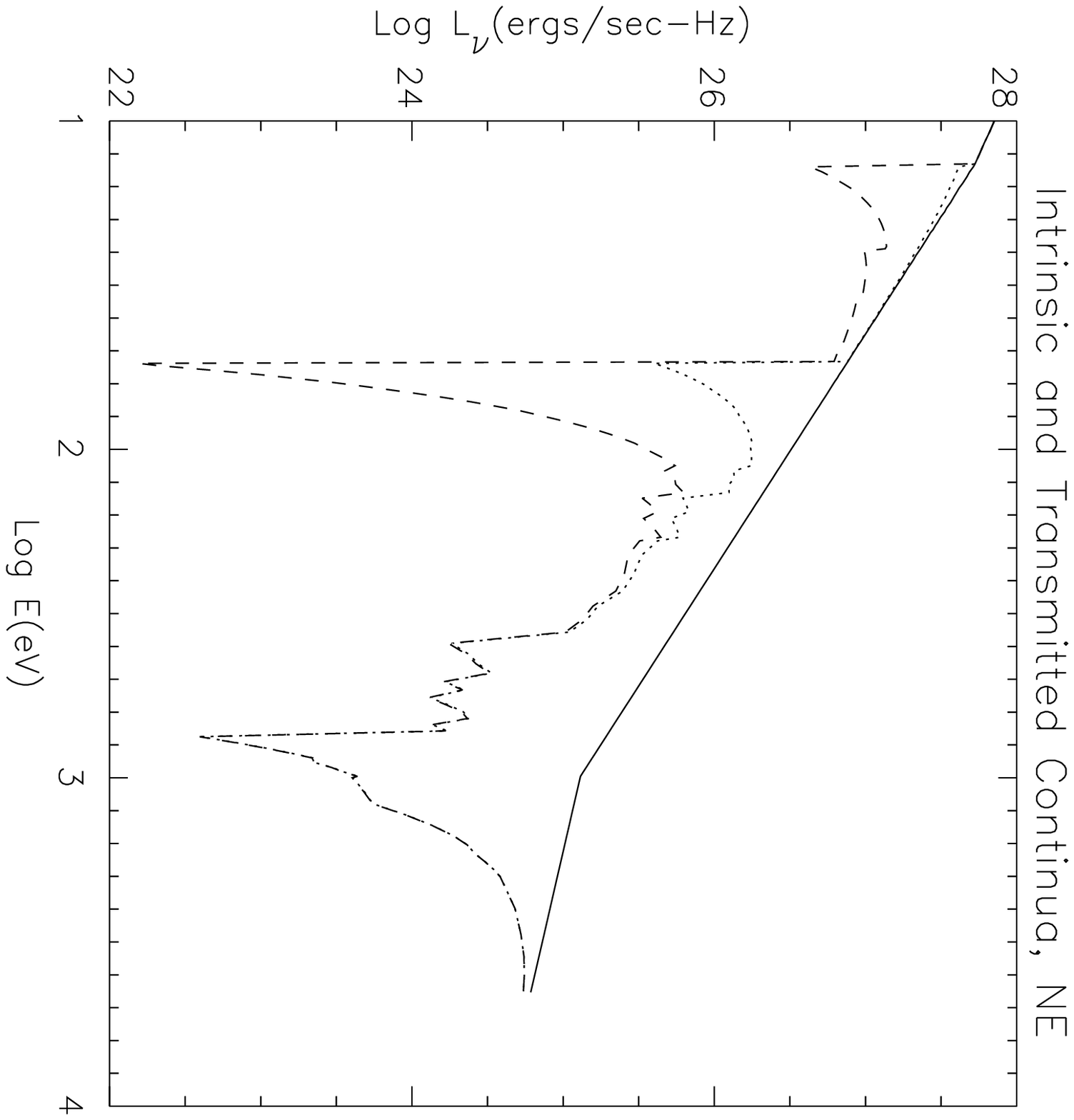}

\clearpage
\plotone{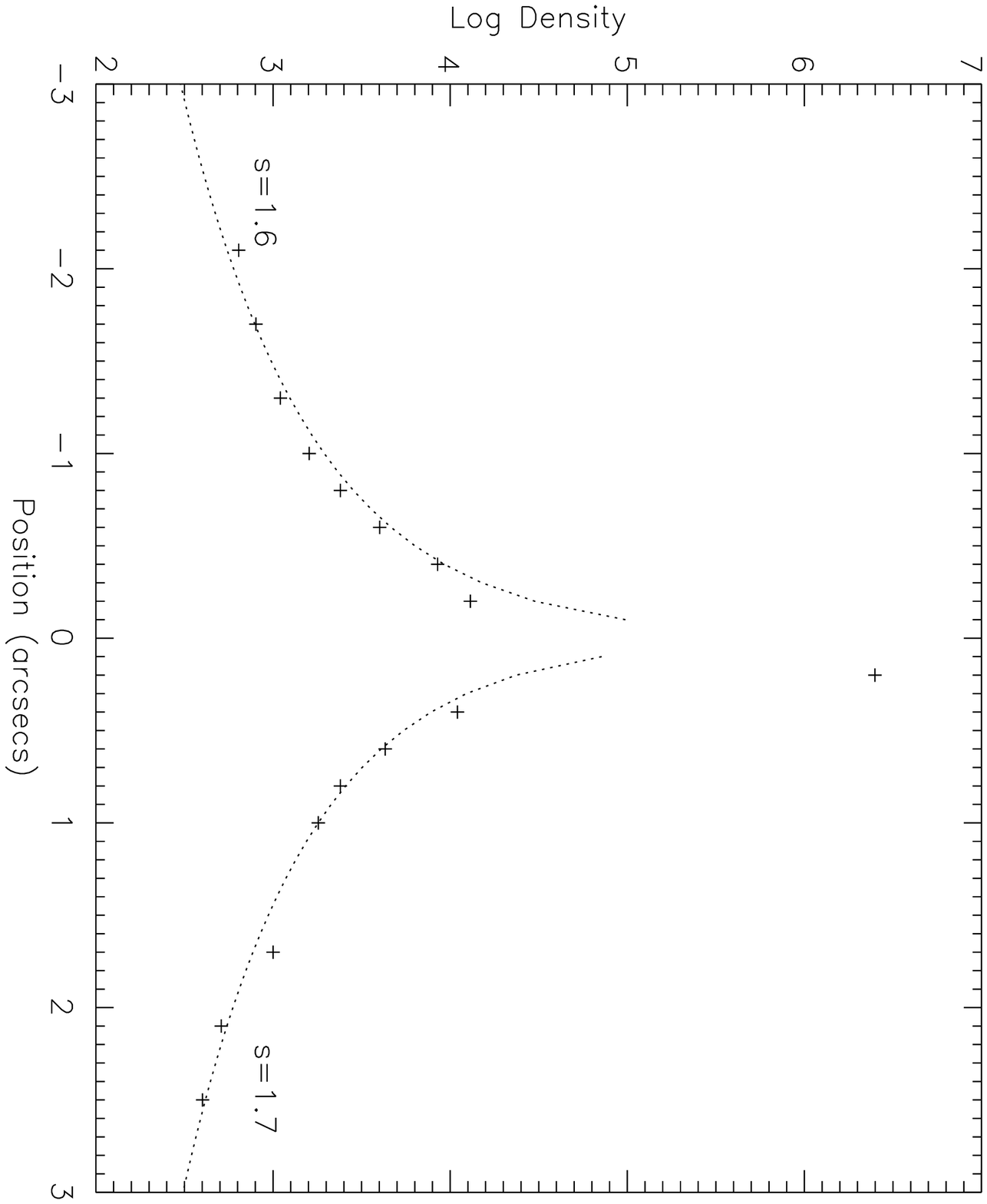}

\clearpage
\plotone{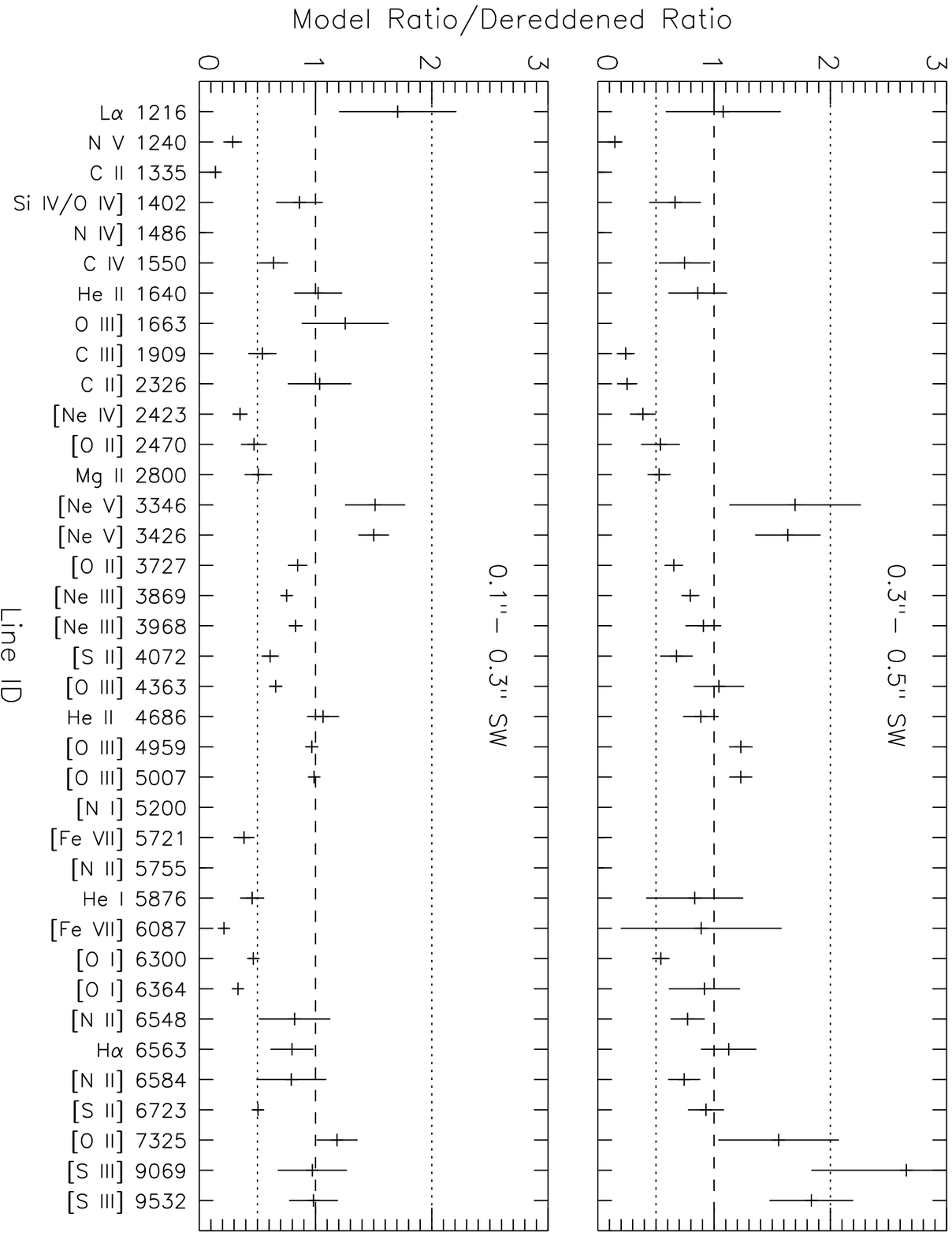}

\clearpage
\plotone{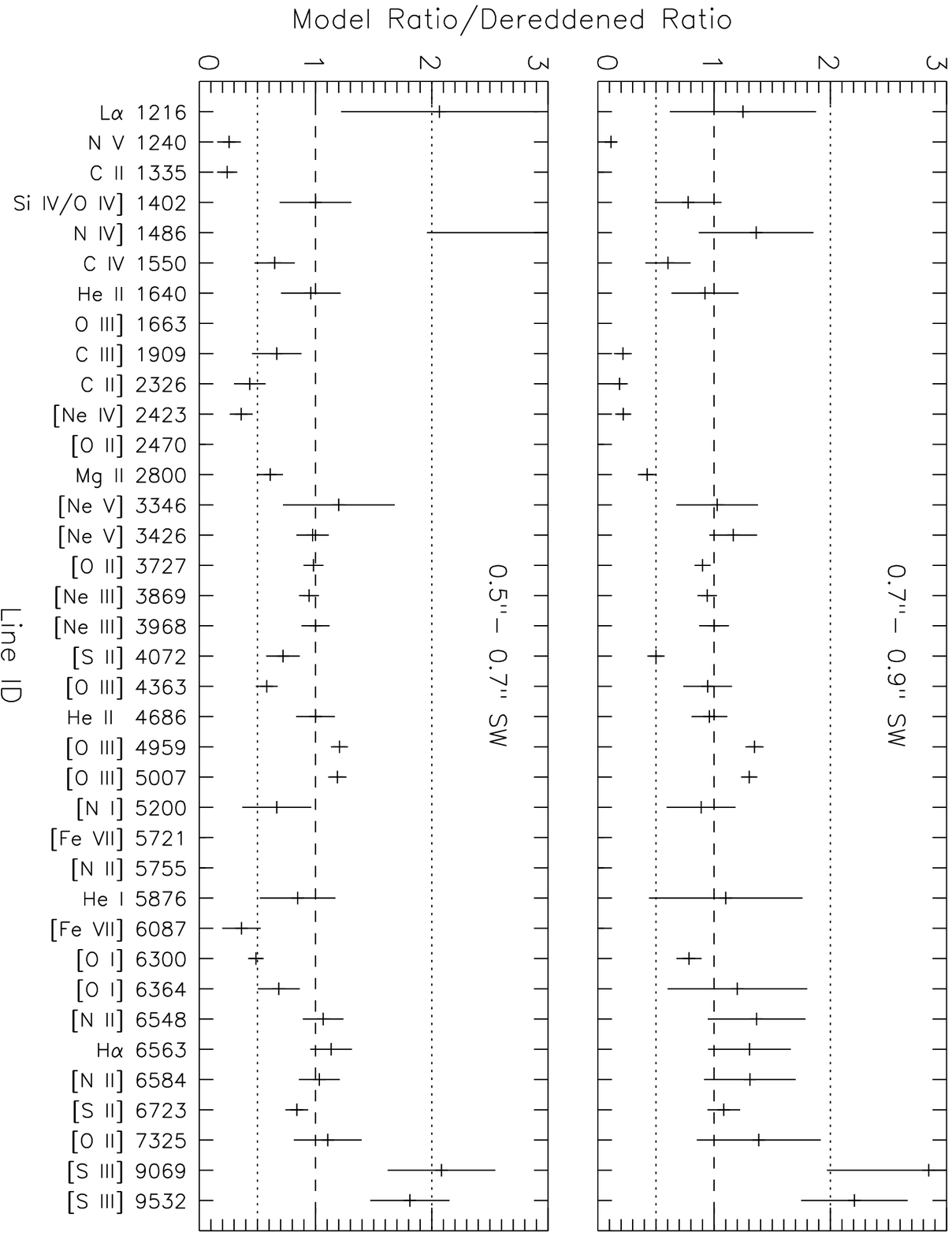}

\clearpage
\plotone{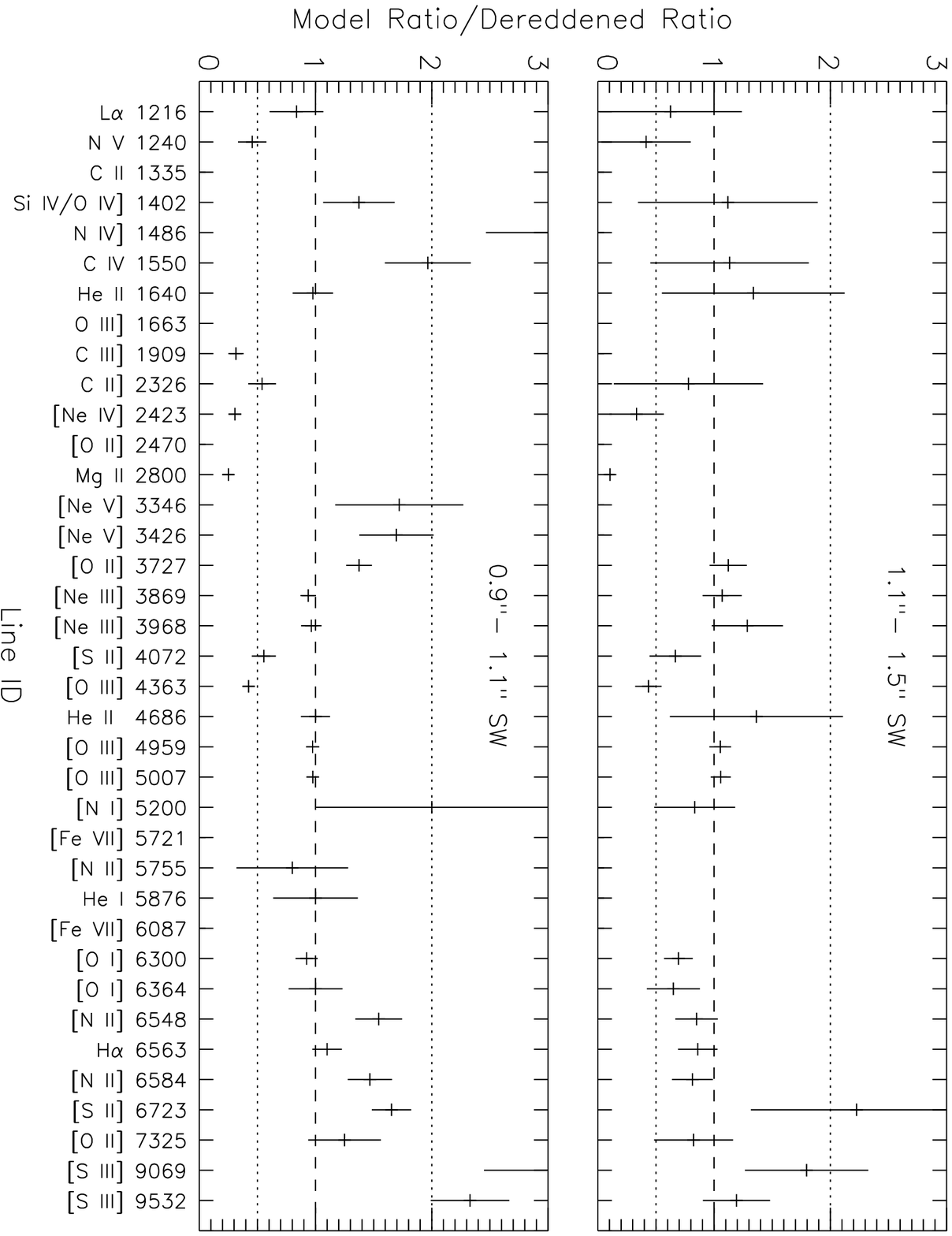}

\clearpage
\plotone{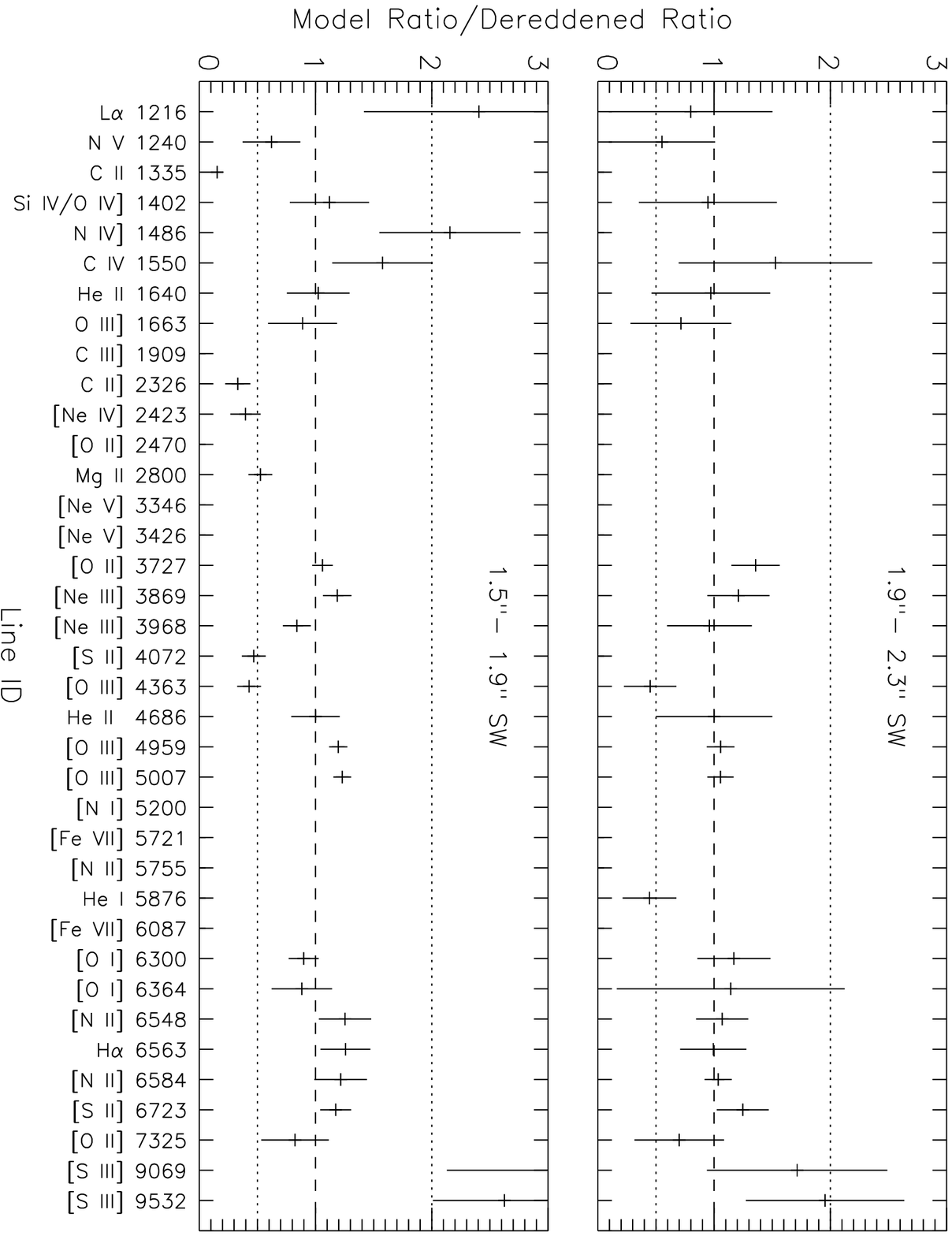}

\clearpage
\plotone{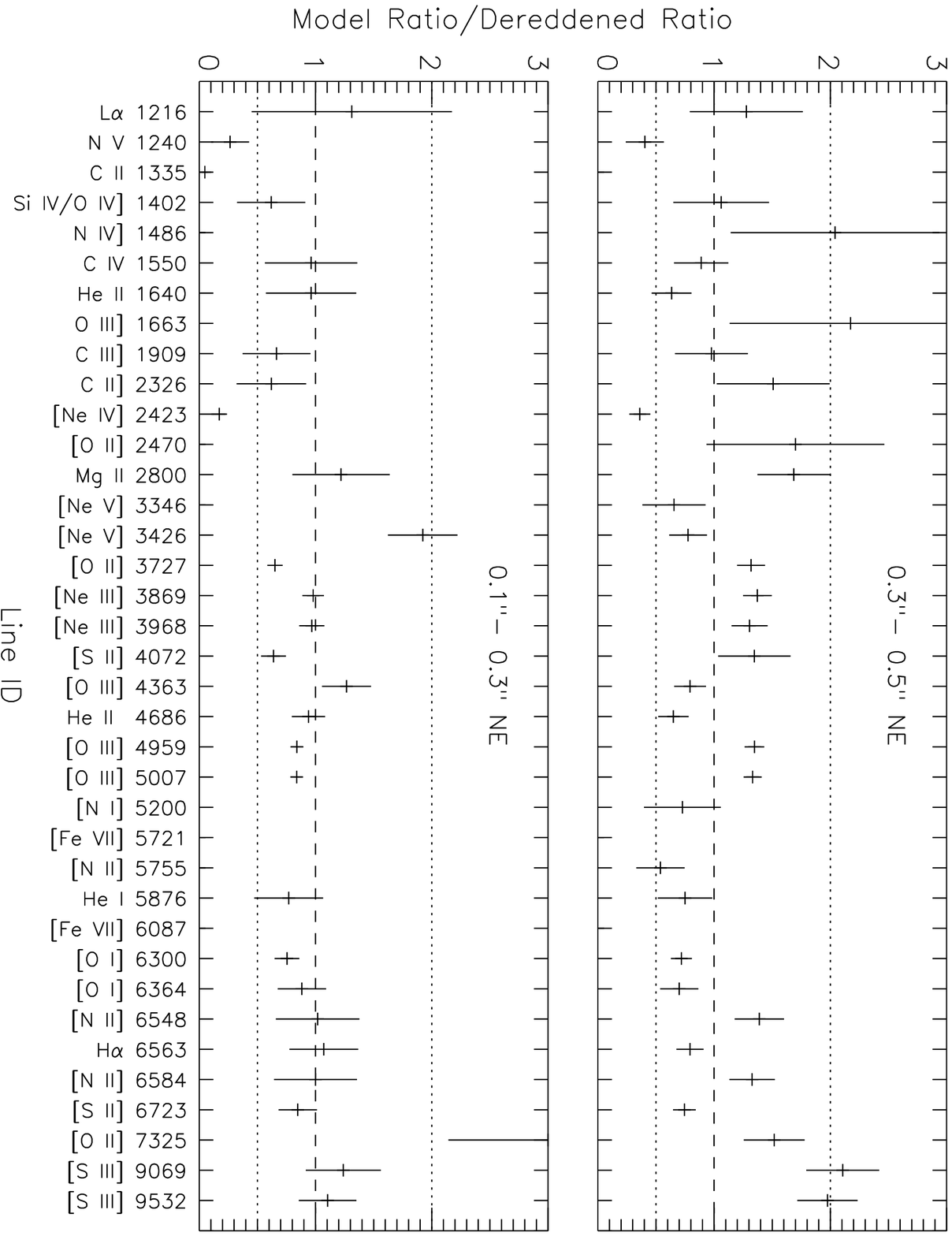}

\clearpage
\plotone{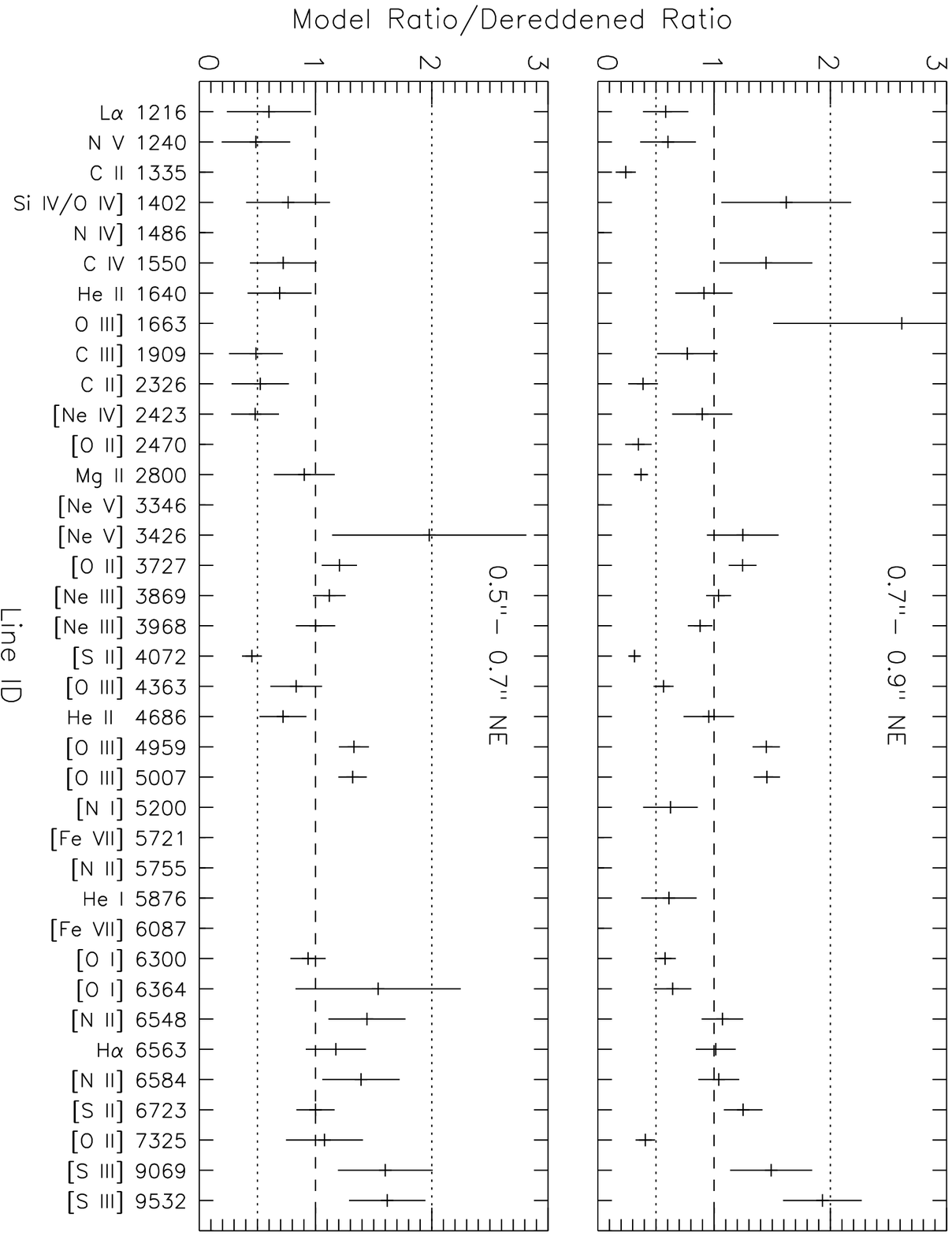}

\clearpage
\plotone{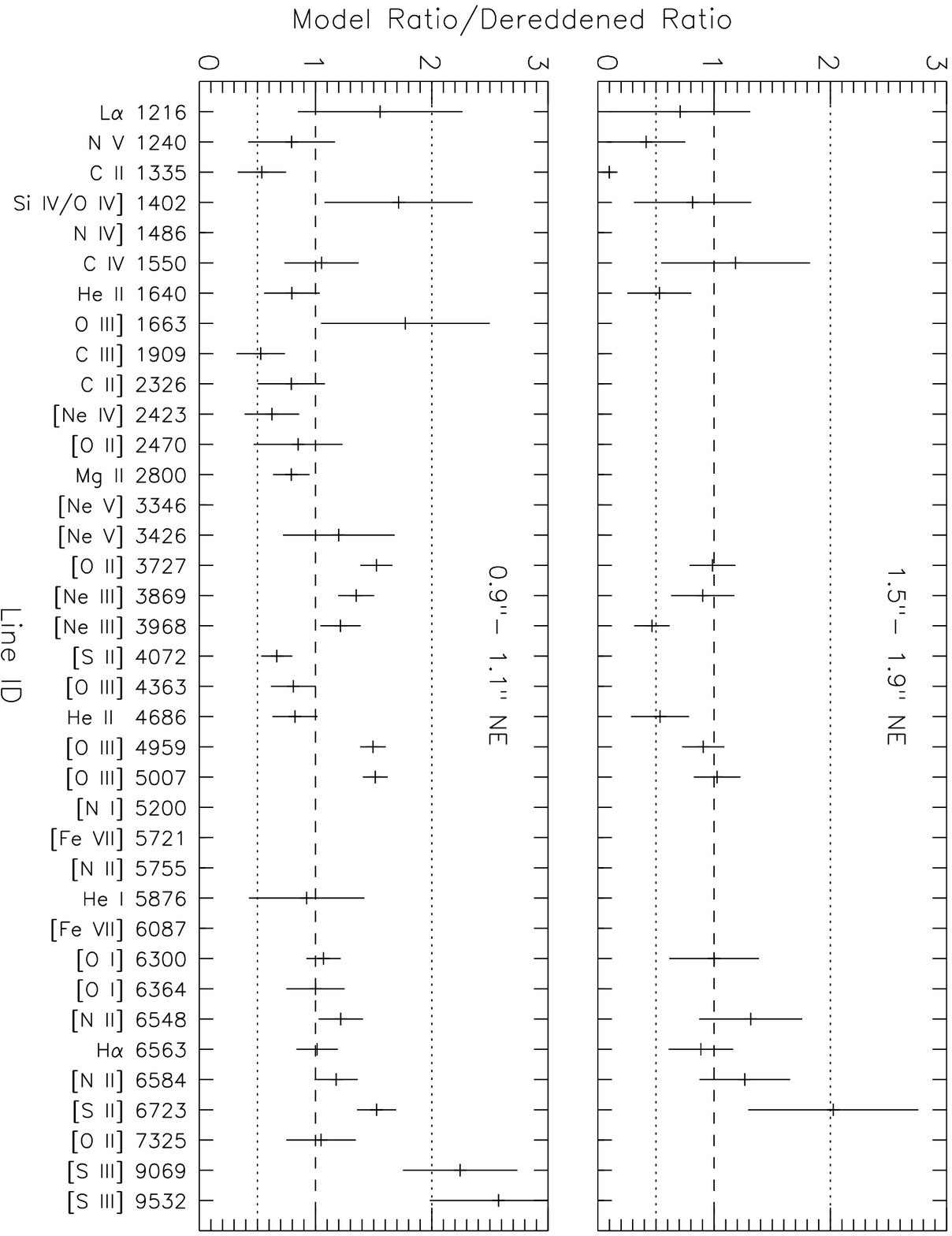}

\clearpage
\plotone{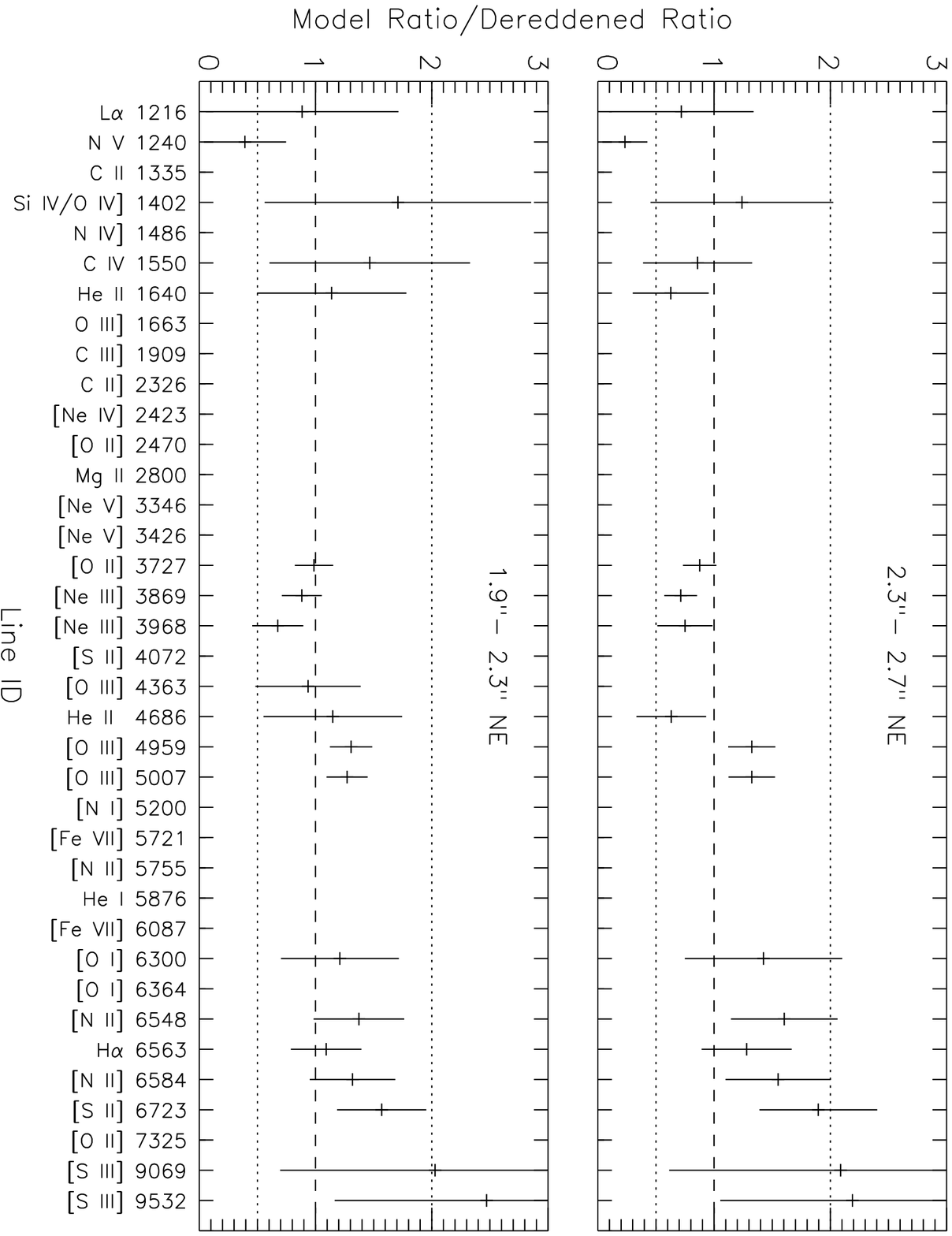}

\clearpage
\plotone{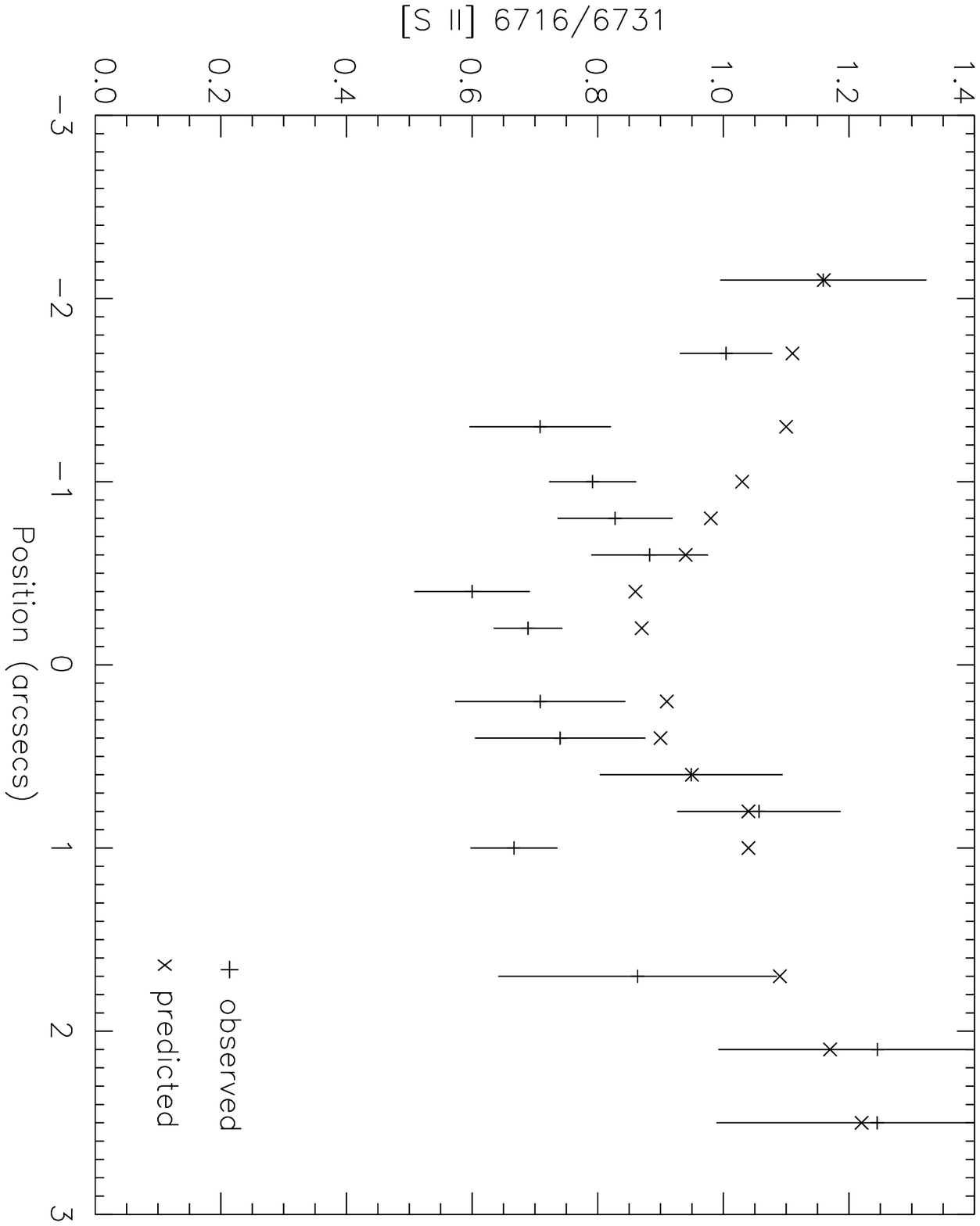}

\end{document}